\documentclass[journal]{IEEEtran}

\usepackage{amsmath,amssymb,amsfonts}

\usepackage[caption=false,font=normalsize,labelfont=sf,textfont=sf]{subfig}
\usepackage{stfloats}
\usepackage{verbatim}
\usepackage{graphicx}

\usepackage{bm}
\usepackage{cite}
\usepackage[colorlinks]{hyperref}

\usepackage{algorithm}
\usepackage{algorithmic}

\begin{document}

% \title{Incentivizing PoS Blockchain for Secured Data Collection UAV-Aided IoT: A MADDPG Approach}
% \title{Incentivizing Proof-of-Stake Blockchain Towards Secured Data Collection in UAV-Assisted IoT:\\A Multi-Agent DRL Approach}
\title{Incentivizing Proof-of-Stake Blockchain for\\Secured Data Collection in UAV-Assisted IoT:\\A Multi-Agent Reinforcement Learning Approach}
% \title{Incentivizing Proof-of-Stake-Based Blockchain for Secured Data Collection in UAV-Assisted IoT: A MADDPG Approach}

\author{Xiao Tang, %~\IEEEmembership{Member,~IEEE,}
        Xunqiang Lan, %~\IEEEmembership{Student Member,~IEEE,}
        Lixin Li, %~\IEEEmembership{Member,~IEEE,}\\
        Yan Zhang, %~\IEEEmembership{Fellow,~IEEE,}
        and Zhu Han %~\IEEEmembership{Fellow,~IEEE}
        % <-this % stops a space
% \thanks{The research work reported in this paper is supported by the National Natural Science Foundation of China under Grant 61901378.}% <-this % stops a space
% \thanks{X. Tang is with the School of Electronics and Information, Northwestern Polytechinical University, Xi'an 710072, China, and also with the National Mobile Communications Research Laboratory, Southeast University, Nanjing 210096, China. (e-mail: tangxiao@nwpu.edu.cn)}%
\thanks{X. Tang, X. Lan and L. Li are with the School of Electronics and Information, Northwestern Polytechnical University, Xi'an 710072, China. (email: tangxiao@nwpu.edu.cn, lanxunqiang@mail.nwpu.edu.cn, lilixin@nwpu.edu.cn)}% <-this % stops a space
\thanks{Y. Zhang is with the Department of Informatics, University of Oslo, 0316 Oslo, Norway. (email: yanzhang@ieee.org)}% <-this % stops a space
\thanks{Z. Han is with the Department of Electrical and Computer Engineering at the University of Houston, Houston, TX 77004 USA, and also with the Department of Computer Science and Engineering, Kyung Hee University, Seoul 446-701, South Korea. (email: hanzhu22@gmail.com)}% <-this % stops a space
}
%\thanks{Z. Han is with the Department of Electrical and Computer Engineering, University of Houston, Houston, TX 77004 USA. (email: zhan2@uh.edu)}% <-this % stops a space
%}

% The paper headers
% \markboth{Journal of \LaTeX\ Class Files,~Vol.~14, No.~8, August~2021}%
% {Shell \MakeLowercase{\textit{et al.}}: A Sample Article Using IEEEtran.cls for IEEE Journals}

% \IEEEpubid{0000--0000/00\$00.00~\copyright~2021 IEEE}
% Remember, if you use this you must call \IEEEpubidadjcol in the second
% column for its text to clear the IEEEpubid mark.

\maketitle

\begin{abstract}
% Internet of Things (IoT) can be conveniently deployed while empowering various applications, where the IoT nodes can form clusters to finish certain missions collectively. In this paper, we propose to employ unmanned aerial vehicles (UAVs) to assist the IoT data collection with blockchain-based security provisioning. In particular, the UAVs generate blocks based on the collected data from IoT clusters, and audit the blocks through the proof-of-stake (PoS) consensus mechanism in the UAV-based blockchain network. Towards efficient blockchain operations, a stake pool is constructed at the active UAV which allows investment-based profit sharing with other UAVs. Therefore, we jointly consider the IoT cluster transmissions, stake pool construction, and UAV deployment with distributed treatment to reach the maximum blockchain system throughput. The problem is decoupled into two layers, where the inner layer tackle the IoT transmissions with large-system analysis while the pool operations with one-leader-multi-follower Stackelberg game and the outer layer finds the UAV deployment through multi-agent deep deterministic policy gradient (MADDPG) approach. Through simulation results, we show the convergence the learning process and the UAV deployment, also demonstrated is the performance superiority of our proposal in terms blockchain throughput as compared with the baselines.
The Internet of Things (IoT) can be conveniently deployed while empowering various applications, where the IoT nodes can form clusters to finish certain missions collectively. In this paper, we propose to employ unmanned aerial vehicles (UAVs) to assist the clustered IoT data collection with blockchain-based security provisioning.
In particular, the UAVs generate candidate blocks based on the collected data, which are then audited through a lightweight proof-of-stake consensus mechanism within the UAV-based blockchain network.
To motivate efficient blockchain while reducing the operational cost, a stake pool is constructed at the active UAV while encouraging stake investment from other UAVs with profit sharing. 
The problem is formulated to maximize the overall profit through the blockchain system in unit time by jointly investigating the IoT transmission, incentives through investment and profit sharing, and UAV deployment strategies.
% The system utility is defined as the expected profit through data collection and investment in unit time by considering the overall blockchain operations. Correspondingly, the problem is formulated as system utility optimization by jointly investigating the IoT transmission, incentive through investment and profit sharing, and UAV deployment strategies.
Then, the problem is solved in a distributed manner while being decoupled into two layers. The inner layer incorporates IoT transmission and incentive design, which are tackled with large-system approximation and one-leader-multi-follower Stackelberg game analysis, respectively. The outer layer for UAV deployment is undertaken with a multi-agent deep deterministic policy gradient approach.
Results show the convergence of the proposed learning process and the UAV deployment, and also demonstrated is the performance superiority of our proposal as compared with the baselines.
\end{abstract}


\section{Introduction}

\IEEEPARstart{I}{nternet} of Things (IoT) is the enabling technology for ubiquitous sensing, computation, and communication towards future wireless networks~\cite{6g}. IoT devices usually feature low-cost, low-power operations with massive and convenient deployment, thus empowering different applications ranging from agriculture, industry, city management, and beyond~\cite{IoTSurvay}. Towards this vision, the data generated by the massive IoT nodes provides the fundamental ingredients, and thus efficient and secure data collection and processing are of significant importance. However, as IoT devices can be deployed in a wide area, the data collection can be quite challenging, and the data processing is usually beyond the local computation capability of IoT devices~\cite{data}. Further, the direct data feedback to the core network is difficult due to the limited resources of IoT devices, and data transmissions can be endangered due to various security threats~\cite{xiao}.

With the rapid development of unmanned aerial vehicle (UAV) technology in recent years, UAVs have been playing an increasingly important role in wireless communications to extend network coverage to three-dimensional space and larger areas~\cite{UAVSurvey}. Attracted by the wide and flexible applications of UAVs, we can dispatch UAVs to reach the vicinity of IoT devices and establish communications therein without conventional network infrastructures, enabling UAV-assisted IoT. With the help of UAVs, an IoT network can be conveniently extended to remote areas with diverse applications. The high mobility of UAVs with flexible deployment and flying provides a new dimension for network optimization to enhance the performance~\cite{UAVIoT}. In this regard, with UAV-assisted IoT, not only the capital and operational expenditures can be saved, but also the efficiency and performance of various IoT applications can be improved.

Despite the facilitation by UAVs in IoT operations, additional strategies are required to address the security issues, since there are various attacks that the defense may be beyond the capability-limited IoT devices. Moreover, the distributed nature of IoT hinders conventional centralized security management relying on network infrastructure~\cite{IoTSec}. Towards this issue, blockchain technology has emerged as a radical solution that provides a transparent, cryptographic, and immutable data structure~\cite{xiong}. Blockchain has dispersed its application and momentum in various areas and brought paradigm shifts therein~\cite{bc_survey}. Meanwhile, Due to the decentralized nature of blockchain, applications of blockchain in the context of IoT have been recognized as an effective solution for security enhancement, identity protection, privacy, and trust management. Supported by blockchain technology, various IoT data can be immutably recorded and encapsulated into blocks to be shared and synchronized among all participants in a distributed manner~\cite{bc_iot_survey}.

Fascinated by the advantages of blockchain technology, it is expected to provide an effective solution to the security provisioning in UAV-assisted IoT. However, the conventional blockchain with proof-of-work (PoW) consensus is computation-intensive and storage-demanding, whose requirements may be beyond the resource-limited IoT devices and dynamic IoT scenarios. In this respect, an external computing server is usually leveraged to assist blockchain operations. Alternatively, we can resort to the proof-of-stake (PoS) blockchain that features mild cost, sufficient scalability, and short delay in the IoT context. For UAV-assisted IoT data collection, a PoS blockchain can be established at the aerial collectors, allowing decentralized tamper-proof security provisioning while effectively avoiding the single-point failure under conventional centralized approaches. Further, the flexible mobility of UAVs can be exploited as an additional dimension for optimization with joint consideration upon IoT transmissions, in order to reduce the latency in different stages of blockchain operations. However, an extensive literature review indicates that the existing studies on blockchain-secured IoT more often leverage external resources for PoW or migrate PoS directly, while lacking consideration and exploitation of the network dynamics. Therefore, there is an urgent need for a lightweight and efficient blockchain solution specially designed for UAV-IoT scenarios for secured data collection.

Towards the aforementioned issues, we propose a UAV-assisted IoT data collection scheme that is secured through PoS blockchain with incentive design. In particular, we consider a clustered IoT with a UAV collecting data for each cluster. The UAVs constitute a blockchain network with block generation from collected data and block audition with the PoS consensus mechanism. In particular, the main contributions can be summarized as follows:
\begin{itemize}
  % \item We propose a UAV-assisted IoT data collection scheme with blockchain, where the amount of blockchained IoT data collected along with the latency incurred at different stages are modeled. We jointly consider the IoT transmissions, incentive design for PoS consensus, and UAV deployment to achieve the highest system throughput-based utility.
  \item For the proposed PoS blockchain-secured UAV-IoT data collection, we consider the obtained profit in the blockchain system through block generation and stake investment, as well as the time consumed for all stages in blockchain operations, and formulate the problem to maximize the system utility as the achieved profit in unit time.
  % \item For the considered scenario, the UAVs as blockchain users obtain profit through block generation and stake investment. 
  % \item For the blockchain-secured UAV-assisted IoT data collection, we consider the profit gaining through blockchain system with IoT data as well as the time consumed for all stages in blockchain operations, and formulate the problem to maximize the system utility as the achieved profit in unit time.
  \item For each UAV-IoT cluster pair, the IoT transmission strategy is designed based on large-system analysis. Then, to facilitate the PoS consensus with reduced operational cost, we propose to construct a stake pool at the active UAV, allowing stake investment and profit-sharing with other UAVs. The incentive process is formulated within a one-leader-multi-follower Stackelberg game framework with equilibrium analysis.
  \item At the networked scale, the distributed UAV deployment is investigated as a multi-agent Markov game, where the instantaneous reward is determined based on the obtained utility from the blockchain system, and the deployment is solved through a multi-agent deep deterministic policy gradient (MADDPG) approach.
  % \item We provide extensive simulation results to show the obtained deployment for the UAV-assisted IoT, also demonstrated is the effectiveness in terms of incentive design as well as the performance superiority in terms of achieved system utility.
\end{itemize}

% with blockchained security. In particular, we consider the clustered IoT operations that each cluster has its serving UAV for data forwarding. Then, the UAV with collected data works as a miner to bundle the data into blocks and employ the PoS consensus protocol to propagate the mined blocks to other UAVs, which works as verifiers to confirm the blocks and chain to the ledger. Correspondingly, we formulate the problem that jointly optimizes the IoT transmissions and UAV deployment to maximize the blockchain throughput. The problem is decomposed into two layers, where the inner layer of IoT transmission is derived with closed-form solutions while the outer layer of UAV deployment is approximated to the optimum with a deep deterministic policy gradient (DDPG)-based learning approach. Finally, simulation results show the convergence of the proposed scheme and demonstrate the performance superiority over the baselines.

% The underlying motivation for blockchain-secured mechanism is that, the different UAV-cluster pairs may belong to different operators for different services, and thus the centralized management of the IoT data is not possible due to the individual interest of the operators. Different from the conventional computation-intensive process as proof-of-work that may be resource-demanding for the energy-limited IoT nodes and UAVs, we in this work adopt the PoS-based blockchain technology among the UAV-formed peer-to-peer network to maintain a blockchain for the IoT data.

The rest of this paper is organized as follows. In Sec.~\ref{sec:rw}, we review the related works. In Sec.~\ref{sec:sys}, we introduce the system model of blockchain-secured IoT data collection with UAVs. In Sec.~\ref{sec:prob}, the problem is formulated to optimize the system utility in terms of the obtained profit in unit time. In Sec.~\ref{sec:inner}, the inner problem is solved for the IoT transmission strategy and incentive design. In Sec.~\ref{sec:outer}, the outer problem for UAV deployment is solved with the MADDPG approach. Sec.~\ref{sec:sim} provides the simulation results to demonstrate the performance, and finally Sec.~\ref{sec:con} concludes this paper.

% \newpage

\section{Related Works} \label{sec:rw}

% uav iot || uav iot security

UAV-facilitated data collection naturally appears as a flexible and effective solution for IoT applications and thus has attracted research interests in various topics~\cite{UAVIoT}. In~\cite{UAVIoTdata}, the authors investigate the UAV trajectory and resource management for time-sensitive IoT data collection while maximizing the number of served IoT devices. In~\cite{UAVdep}, the authors optimize the three-dimensional deployment of multiple UAVs with network interference management to minimize the uplink transmit power of IoT devices. In~\cite{cja}, the authors jointly consider the UAV trajectory, IoT transmission, and scheduling towards energy-efficient data collection. Meanwhile, security guaranteeing rises as a fundamental issue in UAV-IoT scenarios, and also has been investigated in different aspects ranging from physical-layer secrecy to upper-layer cryptography. In~\cite{uaviotphysec}, the authors propose to safeguard the UAV-IoT communications in the physical layer, where the secrecy performance is investigated in the presence of randomly located eavesdroppers with stochastic geometry-based analysis. In~\cite{uaviotaccess}, the authors propose an access control strategy for UAV-assisted IoT for environment surveillance. In~\cite{uaviottrust}, the authors propose a trust evaluation model for IoT data collection, where the UAV-collected data is cleaned to avoid malicious mobile collectors. In~\cite{uaviotprivacy}, the authors consider the UAV-assisted federated learning with the incentive-compatible contract design to protect the privacy of IoT devices. The work above suggests that UAVs can be exploited as flexible yet powerful roles to improve the performance of IoT in various aspects, laying the foundation for our proposed UAV-facilitated IoT security in this work.

% iot bc || uav iot bc

Since the decentralized operation of blockchain naturally fits the IoT scenarios, there have emerged many recent works that apply blockchain in various aspects in the context of IoT. In~\cite{storage}, the authors employ a blockchain-enabled distributed data storage scheme for IoT, where the mining process is exploited for transaction verification as an alternative for the conventional centralized server. In~\cite{computing}, the authors propose an untrusted mobile edge computing PoW scheme for a blockchained IoT system, with fair computing resource allocation among the IoT nodes. In~\cite{pri}, the authors propose a game-based pricing solution between the computation-sensitive node and the cloud server to reach the consensus.
% In~\cite{energy}, the authors develop an energy-efficient dynamic task offloading algorithm for mobile edge-assisted IoT with blockchain-secured data integrity.
However, for the aforementioned works with the public blockchain architecture, they may not be readily extended to UAV-assisted IoT due to the resource-demanding PoW consensus mechanism. In contrast, lightweight solutions requiring relatively lower computing capability and smaller storage are propounded to fit the resource-constrained IoT systems~\cite{cbc}. In~\cite{X2}, the authors propose a soft security scheme for PoS Internet of vehicles based on reputation and contract design. In~\cite{trade}, the authors propose a trading model that allows UAVs to conduct blockchain operations for IoT data in exchange for coins to get recharged. In~\cite{vir}, the authors propose a UAV virtualization scheme with a partially decentralized blockchain model to secure industry IoT on a pay-per-use basis. In~\cite{pos_price}, the authors employ contract theory to balance the stakes and efforts in blockchain IoT towards the maximum profit while tackling the practical scenarios with hidden information and hidden action. In~\cite{consortium}, the authors apply the consortium blockchain with delegated PoS consensus to achieve traceable and anonymous vehicular IoT. In~\cite{pos_uav}, the authors propose a drone-based delegated PoS for IoT to enhance decentralized security with reduced latency. These studies intend for specially designed blockchain systems to fit the lightweight IoT network, the effort is mostly devoted to the consensus mechanism. While in UAV-assisted IoT, the UAV-facilitated dynamics can be actively exploited to improve the performance of blockchain, which is seldom addressed and thus deserves further investigation.

% learning iot bc

Meanwhile, the rapid development of artificial intelligence has advocated the learning-based solution in the wireless area~\cite{S1}. In~\cite{l_bcdrl}, the authors combine deep reinforcement learning and blockchain techniques for IoT data collection, where the former is for the highest throughput while the latter is for security. In~\cite{S2}, the authors propose to exploit deep reinforcement learning to fog network optimization to support IoT. In~\cite{l_bcdl}, the authors propose to exploit blockchain in the deep learning operations in the IoT system to safeguard the learning procedure. In~\cite{l_bct}, the authors propose a privacy-preserving framework for the cooperative Internet of vehicles with blockchain-secured data and deep learning-based prediction. In~\cite{l_bcs}, the authors design a deep \textit{Q}-network-based sharded blockchain for massive IoT services, where the shards improve the system scalability and the deep network finds the optimal throughput configuration. In~\cite{l_bcf}, the authors propose to employ federated learning to protect the IoT data privacy while the learning process is integrated into the consensus process of permissioned blockchain.
% In~\cite{l_bcp}, the authors investigate the space-air-ground integrated power IoT with blockchain-based long-term security and learning-powered queuing delay minimization.
In~\cite{l_bca}, the authors address the priced resource sharing in IoT with blockchain tasks and UAV-based edge computing, and the formulated stochastic game is tackled by hierarchical deep learning techniques. The work above demonstrates the effectiveness of learning techniques in wireless applications, inspiring us to jointly exploit the conventional optimization as well as learning approaches to reach an efficient solution in the UAV-IoT context.

% For example,  Moreover, as UAVs from different operators may be used in IoT, the blockchain technology can be adopted to bundle the collected data into chained blocks towards secure and immutable ledgers with decentralized operations~\cite{bc}. In this respect, the proof-of-stake (PoS)-based blockchain that features mild cost, sufficient scalability, and short delay becomes an attractive solution in the resource-constrained UAV-IoT networks~\cite{cbc}. For example, 

\begin{figure*}[t]
  \centering
  \includegraphics[width=18cm]{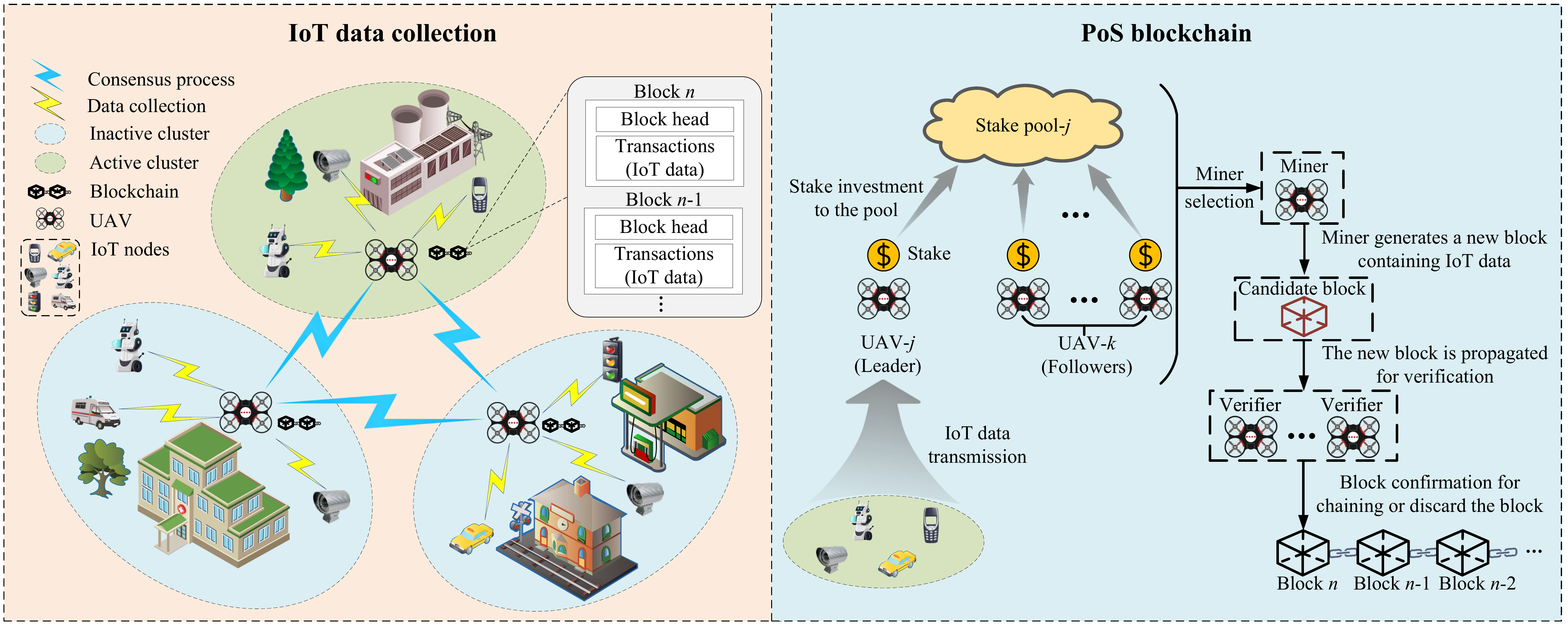}
  \caption{System model.}
  \label{fig:sys}
\end{figure*}

\section{System Model} \label{sec:sys}

We consider a IoT network located within an area denoted by $ \Lambda $. The IoT devices form clusters for certain missions, where the single-point IoT device can be regarded as a special case for the clustered operations. Here, the IoT clusters may belong to different owners and thus work independently. Consider that there are $ J $ clusters, denoted as $ \mathcal{J} = \left\{1,2, \cdots, J\right\} $, and for notation simplicity, we use $ \mathcal{J}_{-j}$ to denote all elements in $ \mathcal{J} $ other than $j$, i.e., $ \mathcal{J}\backslash\{j\} $. For the $j$-th cluster, there are $ I_j $ IoT nodes, denoted as $ \mathcal{I}_j = \left\{1,2,\cdots, I_j\right\} $. For the $i$-th node in the $j$-th cluster with $ i\in\mathcal{I}_j $ and $ j\in\mathcal{J} $, it coordinates within $\Lambda$ are denoted by $ \bm{w}_{ji} = \left[w_{ji}^{(x)}, w_{ji}^{(y)}\right] $. To facilitate the IoT operations, one rotary-wing UAV is deployed for each cluster to collect the data, where the UAV also works as a delegate for further data processing in the blockchain system. The UAVs are assumed to hover at a fixed altitude of $ H $ and the one for the $j$-th cluster is of a horizontal coordinates of $ \bm{v}_j = \left[v_j^{(x)}, v_j^{(y)}\right] $. The system model is illustrated as Fig.~\ref{fig:sys}.

% \begin{figure}[t]
%   \centering
%   \includegraphics[width=9cm]{work_procedure.pdf}
%   \caption{Work procedure of data collection and processing.}
%   \label{fig:pos}
% \end{figure}

\subsection{IoT Transmission Model}

For the considered system, the IoT nodes use single-antenna due to their limited size and capability. Meanwhile, the UAV has $ K $ antennas to enhance the receptions. In this respect, the uplink transmissions from the $j$-th cluster to the serving UAV form an $I_j \times K$ dimensional virtual multiple-input-multiple-output (MIMO), for which the transmission model is given as
\begin{equation}
    \bm{y}_j = \bm{H}_j\bm{x}_j + \bm{z}_j,
\end{equation}
where $ \bm{x}_j $, $ \bm{y}_j $, and $ \bm{z}_j $ are the $ I_j $-dimensional transmitted signal, $ K $-dimensional received signal, and background noise, respectively, and $ \bm{H}_j \in\mathbb{C}^{K\times I_j} $ is the virtual MIMO channel. Based on the large-system analysis technique~\cite{appx}, the channel can be further decomposed as
\begin{equation}
    \bm{H}_j = \bm{S}_j \bm{L}_j,
\end{equation}
where $ \bm{L}_j $ is the large-scale component and $ \bm{S}_j $ corresponds to the small-scale fading. Specifically, $ \bm{L}_j $ is an $ I_j $-dimensional diagonal matrix given as $ \mathsf{diag}\left(\left[\ell_{ji}^{1/2}\right]_{i\in\mathcal{I}_j}\right) $ with $ \ell_{ji}^{1/2} = d_{ji}^{-2}10^{-\delta_{ji}/10} $, where $ d_{ji} $ is the distance between the IoT node and UAV as
\begin{equation}
    d_{ji} = \sqrt{\left\|\bm{w}_{ji} - \bm{v}_{j}\right\|^2 + H^2},
\end{equation}
and
\begin{equation}
  \delta_{ji} = 20\log_{10}\left(\frac{4\pi f}{c}\right) + \frac{\eta_{\mathsf{LoS}} - \eta_{\mathsf{NLoS}}}{1 + a\exp\left(-b({\phi}_{ji}-a)\right)} + \eta_{\mathsf{NLoS}}
\end{equation}
is the combined effect of line-of-sight (LoS) and non-LoS (NLoS) fading with $ f $ and $ c $ being the carrier frequency and speed of light, respectively, $ \eta_{\mathsf{Los}} $, $ \eta_{\mathsf{NLos}} $, $ a $, and $ b $ depending on the propagation environment, and $ {\phi}_{ji} = \frac{180}{\pi}\arcsin\frac{H}{d_{ji}} $.

For the IoT data collection, at each time instant there is only one active cluster, and thus the data collection process is inter-cluster interference-free. This corresponds to the practical scenarios where IoT data is in small amount and thus the transmission can finish rather quickly. For node-$i$ in cluster-$j$, the transmit power is $ p_{ji} $ and limited by the power constraint given by
\begin{equation} \label{eq:ind_pwr_con}
    0 \le p_{ji} \le p_{j}^{\text{msk}}, \quad \forall i \in\mathcal{I}_j,
\end{equation}
and
\begin{equation} \label{eq:sum_pwr_con}
    \sum_{i\in\mathcal{I}_j} p_{ji} \le p_j^{\text{max}},
\end{equation}
where $ p_{j}^{\text{msk}} $ and $ p_j^{\text{max}} $ are the per-node and per-cluster maximum power, respectively.
Then, the transmission rate in cluster-$j$ is obtained as
\begin{equation} \label{eq:rate}
  r_{j}^{\text{(ag)}} = \mathbb{E}_{\bm{S}_{j}} \left\{ B^{\text{(ag)}} \log_2\det\left( \bm{I}_K + \frac{\bm{S}_{j}\bm{L}_{j}\bm{P}_{j} \bm{L}_{j}^T\bm{S}_{j}^T}{\sigma_0^2} \right) \right\},
\end{equation}
where $ B^{\text{(ag)}} $ is the bandwidth with superscript indicating air-ground transmissions, $ \bm{P}_{j} = \mathsf{diag}\left( [{p}_{ji}]_{i\in\mathcal{I}_j} \right) $ collects the transmit power of all nodes in cluster-$j$, $ \bm{I}_K $ is a $K$-dimensional identity matrix, $ \sigma_0^2 $ is the background noise power, and the expectation is conducted with respect to small-scale fading.

\subsection{Blockchain Model}

For the UAV-assisted IoT model, we consider that trust has been established in each cluster between the IoT nodes and UAV, yet they have no information regarding the legitimacy of other clusters. Thus, to defend against the data tampering by the potential malicious adversaries, blockchain is introduced to process the collected IoT data to guarantee the security and integrity in a decentralized manner. The blockchain is established among the UAVs, where the PoS consensus mechanism is employed with mild operational cost while being convenient to incorporate the incentive designs. As a complete model for PoS blockchain operations can be rather cumbersome, and thus we concentrate on the principle components incurring the latency to reach the consensus, including transmission, mining, propagation, verification, and confirmation. Assume the data amount collected from cluster-$j$ is denoted by $ \Psi_j $ (in bit), the following time components need to be consumed for the data to be bundled, audited, and chained to the existing ledger.

\textit{1) Transmission time:}
Given the transmission model introduced before for the IoT uplink in cluster-$j$, the time consumed for $ \Psi_j $ bits to reach the serving UAV is
\begin{equation} \label{eq:t_tx}
  \tau_j^{(\text{tx})} = \frac{\Psi_j}{r_{j}^{\text{(ag)}}},
\end{equation}
where $ r_{j}^{\text{(ag)}} $ is obtained from~(\ref{eq:rate}).

\textit{2) Mining time:}
When the UAV finishes the data collection from the served IoT cluster, the mining process is conducted to generate candidate blocks with data regarded as transactions. The time consumed in this part depends on the computation capability of the UAV, given as
\begin{equation}
  \tau_j^{(\text{mn})} = \frac{\Psi_j}{\zeta}\nu,
\end{equation}
for cluster-$j$, where $ \zeta $ is the computation rate (in bit/s), and $ \nu $ is the coefficient of computation complexity for mining. Note that the data size for mining is slightly larger than $ \Psi_j $ considering the required block header. But for a general case, the IoT data (transactions) dominates the overall size and we can safely use $ \Psi_j $ as the mining data size. This assumption is also applied for propagation and verification time modelings. As we adopt the PoS consensus mechanism, the mining complexity can be largely alleviated.

\textit{3) Time for propagation and verification:}
For the mined blocks by the current UAV, they will be propagated to other UAVs in the blockchain network for verification and only the verified blocks can be chained. In this process, the mining UAV broadcasts the candidate blocks, for which the propagation delay is given as
\begin{equation}
  \tau_j^{(\text{pv})} = \frac{\Psi_j}{\min\limits_{j,k\in\mathcal{J}} r_{j,k}^{(\text{aa})}},
\end{equation}
where $ r_{j,k}^{(\text{aa})} $ is the transmission rate from UAV-$j$ to UAV-$k$ with superscript indicating air-to-air links. Evidently, the propagation delay depends on the inter-UAV communications with the lowest transmission rate. While the inter-UAV communication experiences LoS air-to-air propagation model and thus the rate from UAV-$j$ to UAV-$k$ can be defined as
\begin{equation}
  r_{j,k}^{(\text{aa})} = B^{(\text{aa})}\log_2\left( 1 + \frac{K P_j \left\| \bm{v}_j - \bm{v}_k \right\|^{-2} }{\sigma_0^2} \right),
\end{equation}
where $ B^{(\text{aa})} $ is the bandwidth, $ P_j $ is the transmit power of UAV-$j$, $ K $ arises as the array gain at the receiving UAV with $ K $ antennas. Here, the interference is not explicitly incorporated regarding the inter-UAV communications due to the asynchronized blockchain operations among the UAVs. Then, the propagated candidate block is verified at other UAVs. As the verification corresponds to certain hash calculation that can be done rather quickly, the time consumed can be neglected as compared with propagation time, and thus is not explicitly considered here.

\textit{4) Time for confirmation and chaining:}
When the block verification is finished, the results are fed back to confirm that the block is valid through the air-to-air links among the UAVs. The candidate block is then chained into the current ledger and the local version of the ledger at all client updates correspondingly. This process includes some lightweight transmission and computation, the time required is denoted by $ \tau_{j}^{(\text{cc})} $, which can be assumed to be a constant.

\section{Problem Formulation and Decomposition} \label{sec:prob}

Given the basic models regarding the IoT data transmission and blockchain operation, we utilize an efficient PoS procedure with incentive design. According to the PoS procedure, in each round there will be a leader, say UAV-$k\in\mathcal{J}$, to conduct the blockchain operation for the IoT data in cluster-$j$ that is currently active. In this process, an amount of $ \Psi_j $ IoT data induces an overall payment of $ \Phi + \rho\Psi_j $ (in coin), where $ \Phi $ is the fixed payment for each valid block and $ \rho $ is the coefficient (in coin/bit) for transactions fees. Meanwhile, the blockchain operation with respect to the data in cluster-$j$ incurs a cost, denoted by $ \Omega_{j,k}$, when the leader processing the blockchain is UAV-$k$. Note that the IoT data originated from cluster-$j$ can be bundled into blocks by any UAV, yet there require additional procedures for UAV-$j$ as the collector to relay the data to UAV-$k$ for blockchaining, and thus we assume that $ \Omega_{j,j} < \Omega_{j,k} $ for all $k\in\mathcal{J}_{-j}$.

Moreover, based on the principle of PoS, the probability for a UAV as the stakeholder to generate a block (i.e., mining), is proportional to its amount of stakes. The stake can be the coin deposit allocated by the network operators to facilitate the blockchain operations. Here, we denote the available stake at UAV-$j$ as $ \Upsilon_j $, $j\in\mathcal{J}$. In accordance with the aforementioned payment and cost models, we encourage the UAV serving the currently active cluster to conduct the mining process, which helps reduce the system cost. To this end, we propose to construct a stake pool at the currently active UAV. Without loss of generality, we assume cluster-$j$ is active, and then a pool is constructed at UAV-$j$. Meanwhile, other UAVs also contribute to the pool with part of their stakes. We assume that the ratio of invested stake at UAV-$k$ is $ \alpha_{j,k} \in\left[0,1\right] $, $k\in\mathcal{J}_{-j}$, then the probability for the pool, i.e., UAV-$j$ to generate the block is
\begin{equation}
  q_{j,j} = \frac{\Upsilon_j + \sum\limits_{k\in\mathcal{J}_{-j}}\alpha_{j,k}\Upsilon_k}{\sum\limits_{j\in\mathcal{J}}\Upsilon_j},
\end{equation}
where UAV-$j$ naturally devotes all its stakes in the pool. Then, the probability for other UAVs being the miner is
\begin{equation}
  q_{j,k} = \frac{\left(1-\alpha_{j,k}\right)\Upsilon_k}{\sum\limits_{j\in\mathcal{J}}\Upsilon_j}, \quad\forall k \in \mathcal{J}_{-j}.
\end{equation}

In accordance with the pool construction at UAV-$j$, the investment of other UAVs should be rewarded to incentivize their cooperation. In particular, UAV-$j$ splits a portion of its potential payment for block generation to be shared with its investors, where the returned payment is proportional to the investment. Denote the portion of payment to reward the investors as $ \beta_j \in \left[0,1\right] $ at UAV-$j$. If the pool successfully works as the miner receiving a payment of $\Phi + \rho\Psi_j$, it obtains a profit of $ \left(1-\beta_j\right)\left(\Phi + \rho\Psi_j\right) - \Omega_{j,j} $, corresponding to the preserved payment while taking out the cost. Meanwhile, UAV-$k$ in $ \in \mathcal{J}_{-j} $ obtains a profit of $ \frac{\alpha_{j,k}\Upsilon_k}{\sum\nolimits_{k\in\mathcal{J}_{-j}}\alpha_{j,k}\Upsilon_k}\beta_j\left(\Phi + \rho\Psi_j\right) $, corresponding to the investment-proportional payment at no cost. Also, there are also possibilities that UAV-$k$ in $ \in \mathcal{J}_{-j} $ creates a candidate block. Then, this UAV obtains a profit of $ \left(\Phi + \rho\Psi_j\right) - \Omega_{j,k} $, earning all the payment with the corresponding cost. Therefore, the expected profit for the pool, i.e., UAV-$j$, is given as
% \begin{equation} \label{eq:ujj}
% \begin{aligned}
% u_{j,j} \left( \beta_j, \bm{\alpha}_j \right) =& \frac{\Upsilon_j + \sum\limits_{k\in\mathcal{J}_{-j}}\alpha_{j,k}\Upsilon_k}{\sum\limits_{j\in\mathcal{J}}\Upsilon_j} \\
% &  \times \left[ \left(1-\beta_j\right)\left(\Phi + \rho\Psi_j\right) - \Omega_{j,j} \right],
% \end{aligned}
% \end{equation}
\begin{equation} \label{eq:ujj}
% \begin{aligned}
\Theta_{j,j} = \frac{\Upsilon_j + \sum\limits_{k\in\mathcal{J}_{-j}}\alpha_{j,k}\Upsilon_k}{\sum\limits_{j\in\mathcal{J}}\Upsilon_j}
\left[ \left(1-\beta_j\right)\left(\Phi + \rho\Psi_j\right) - \Omega_{j,j} \right].
% \end{aligned}
\end{equation}
Also, the expected profit for other UAVs is
% \begin{equation} \label{eq:ujk}
% \begin{aligned}
% u_{j,k} \left( \beta_j,\alpha_{j,k},\bm{\alpha}_{j,-k} \right) = \frac{\left(1-\alpha_{j,k}\right)\Upsilon_k}{\sum\limits_{j\in\mathcal{J}}\Upsilon_j} \left[ \left(\Phi + \rho\Psi_j\right) - \Omega_{j,k} \right] \\
% + \frac{\Upsilon_j + \sum\limits_{k\in\mathcal{J}_{-j}}\alpha_{j,k}\Upsilon_k}{\sum\limits_{j\in\mathcal{J}}\Upsilon_j} \frac{\alpha_{j,k}\Upsilon_k}{\sum\nolimits_{k\in\mathcal{J}_{-j}}\alpha_{j,k}\Upsilon_k}\beta_j\left(\Phi + \rho\Psi_j\right),
% \end{aligned}
% \end{equation}
\begin{equation} \label{eq:ujk}
\begin{aligned}
\Theta_{j,k} =& \frac{\left(1-\alpha_{j,k}\right)\Upsilon_k}{\sum\limits_{j\in\mathcal{J}}\Upsilon_j} \left[ \left(\Phi + \rho\Psi_j\right) - \Omega_{j,k} \right] \\
&+ \frac{\Upsilon_j + \sum\limits_{k\in\mathcal{J}_{-j}}\alpha_{j,k}\Upsilon_k}{\sum\limits_{j\in\mathcal{J}}\Upsilon_j} \frac{\alpha_{j,k}\Upsilon_k}{\sum\limits_{k\in\mathcal{J}_{-j}}\alpha_{j,k}\Upsilon_k}\beta_j\left(\Phi + \rho\Psi_j\right), \\
&\qquad\qquad\qquad\qquad\qquad \forall k\in\mathcal{J}_{-j}.
\end{aligned}
\end{equation}
Note that the above discussions are based on the assumption that the IoT data originates from cluster-$j$, and the formulation can be readily extended to the cases when other clusters are active.
% For the profit defined in~(\ref{eq:ujj}) and~(\ref{eq:ujk}), the first subscript indicates that the IoT data collection originates from cluster-$j$, and second subscript specifies the obtained utility at different UAVs.

Based on the modeling of the profit gaining and time consumed over the blockchain operations with respect to IoT data, we can then define the utility achieved by processing the data from cluster-$j$ as
\begin{equation} \label{eq:uj}
  U_j = \frac{\Theta_{j,j} + \sum\limits_{k\in\mathcal{J}_{-j}}\Theta_{j,k}}{\tau_j^{\text{(tx)}} + \tau_j^{\text{(mn)}} + \tau_j^{\text{(pv)}} + \tau_j^{\text{(cc)}}},
\end{equation}
where the profit of all parties including the pool and other followers is incorporated. With the proposed incentive mechanism, the profit is obtained not only from mining as a conventional blockchain but also from stake investment. In this regard, the incentive design introduces an implicit altruistic effect in the blockchain system. Then, the problem is formulated to optimize the overall system utility by jointly considering the IoT transmissions, stake investment and profit sharing, and the UAV deployment, specified as
\begin{IEEEeqnarray}{cl} \hspace{-10pt}
  \IEEEyesnumber\label{eq:problem} \IEEEyessubnumber*
  \max_{\left\{\bm{v}_j,\bm{p}_j, \beta_j, \bm{\alpha}_j \right\}_{j\in\mathcal{J}}} \quad & U = \sum\limits_{j\in\mathcal{J}} U_j  \\
   \rm{s.t.} \quad & \bm{v}_j\in\Lambda, \quad \forall j\in\mathcal{J}, \label{eq:deploy_con} \\
   & (\text{\ref{eq:ind_pwr_con}}), (\text{\ref{eq:sum_pwr_con}}), \quad \forall j\in\mathcal{J}, \label{eq:pwr_con} \\
   & \beta_j \in[0,1], \quad \forall j\in\mathcal{J}, \\
   & \alpha_{j,k} \in[0,1], \: \forall j\in\mathcal{J}, \: \forall k\in\mathcal{J}_{-j},
\end{IEEEeqnarray}
where $ \bm{p}_j = \left[ p_{ji} \right]_{i\in\mathcal{I}_j} $ is the power vector in cluster-$j$ and $ \bm{\alpha}_j = \left[\alpha_{j,k}\right]_{k\in\mathcal{J}_{-j}} $ is the vector of stake investment portion.

For the formulated problem, we can see that the considered factors affect system performance in a coupled and complicated manner. Particularly, the IoT transmissions constitute the basic ingredient for block generation. The incentives with investment and profit-sharing affect the achieved profit. The UAV deployment and network topology influence the time for blockchain operations such as data transmission and block propagation. Then, it can be rather cumbersome to tackle the problem directly. Moreover, as the optimization in~(\ref{eq:problem}) appears in a centralized manner, the corresponding algorithm design violates the decentralized operations of the UAV-assisted IoT and blockchain system. Therefore, we will then decompose the problem to facilitate the distributed solution.

Revisit the formulated problem to maximize the sum utility within a distributed perspective, we first allow each UAV-IoT cluster to determine their own strategies. Then, by analyzing the relation between the optimization variables and the objective, we have the following observations. The IoT transmission and incentive strategies can be investigated at each UAV-IoT-pair basis since transmissions occur within the current pair while the incentive needs to be designed with respect to the pool constructed for the current pair. In contrast, the UAV deployment affects the performance in the networked scale, as the movement of one UAV not only affects the IoT data transmission in its own cluster, but also the block propagation to other clusters. Therefore, we tackle the problem in a distributed manner to allow the individual and independent decision-making at each UAV-IoT cluster pair. Further, the problem at each UAV-IoT cluster pair is decomposed into two layers where the outer layer solves for the UAV deployment while the inner layer for IoT transmissions and incentive designs. The inner problem can be solved independently for each UAV-IoT cluster pair where the transmission optimization corresponds to the minimization of the denominator of the utility function in~(\ref{eq:uj}), while the incentive design maximizes the nominator. The outer problem is solved through a multi-agent reinforcement learning process for individual optimality at each UAV. Correspondingly, the problem in~(\ref{eq:problem}) is then solved in a decentralized manner allowing the individual decision-making at each party to facilitate the implementation.

% Moreover, the inner-layer problem is solved by exploiting large-system analysis and game-based solution, and the outer-layer problem is tackled through multi-agent deep reinforcement learning, as detailed in the following two sections.

\section{Inner Problem Solving for IoT Transmission and Incentive Design} \label{sec:inner}

Based on the problem decomposition introduced before, we consider the inner problem at each UAV-IoT cluster basis to solve for the transmission and incentive strategy, with fixed UAV deployment at the outer layer. The inner problem intends to maximize the individual utility function in~(\ref{eq:uj}), where the transmission strategy minimizes the denominator of the utility and the incentive design tackles the nominator. Thus, these two subproblems can be addressed independently as detailed below.

\subsection{IoT Transmission Strategy}

In accordance with the problem in~(\ref{eq:problem}), the maximum IoT transmission rate corresponds to the minimized transmission time in~(\ref{eq:t_tx}), leading to the maximization of the objective function in~(\ref{eq:problem}). As has been noted, the transmission strategy is independently determined at each IoT cluster, then the problem is formulated to maximize the rate in~(\ref{eq:rate}) with respect to the power constraint in~({\ref{eq:ind_pwr_con}}) and ({\ref{eq:sum_pwr_con}}). For this problem, we first tackle the expectation operation with large-system analysis technique~\cite{appx} and approximate the transmission rate as
\begin{equation}
    \begin{aligned}
    r_{j}^{\text{(ag)}} = & \: B^{\text{(ag)}} \left[ \sum\limits_{i\in\mathcal{I}_j}\log_{2}(1+\frac{1}{\sigma_{0}^{2}}K\ell_{j,i}p_{ji} \omega_{j}^{-1}) \right. \\
    & \left. +\: K\log_{2}(\omega_{j})-K\log_{2}e (1-\omega_{j}^{-1}) \right],
    \label{Rj}
  \end{aligned}
\end{equation}
where $ \omega_j $ is the newly introduced auxiliary variable satisfying
\begin{equation} \label{eq:omega}
  \omega_{j} = 1+\sum\limits_{i=1}^{I_{j}}\frac{\ell_{j,i}p_{ji}}{\sigma_{0}^{2}+K\ell_{j,i}p_{ji}\omega_{j}^{-1}}.
\end{equation}
Then, the IoT transmission problem becomes
\begin{IEEEeqnarray}{cl}
  \IEEEyesnumber\label{eq:prob_pwr} \IEEEyessubnumber*
  \max_{\bm{p}_j, \omega_j} \quad & r_j^{\text{(ag)}}  \\
   \rm{s.t.} \quad & (\text{\ref{eq:ind_pwr_con}}), (\text{\ref{eq:sum_pwr_con}}), \text{ and } (\text{\ref{eq:omega}}),
\end{IEEEeqnarray}
for cluster-$j\in\mathcal{J}$. For this problem, we can adopt the alternating optimization to tackle the power optimization and auxiliaries in an iterative manner. In particular, with fixed auxiliaries, the power allocation is evidently a convex optimization and thus we can leverage the Lagrange multiplier method to obtain that
\begin{equation}
  p_{ji}^{\star} = \left( \mu_j - \frac{\omega_j\sigma_0^2}{K\ell_{ji}} \right)_0^{p_{j}^{\text{msk}}}
\end{equation}
with $ (\:\cdot\:)_a^b $ indicating $ \min(\max(\cdot, a), b) $ and $ \mu_j $ being the multiplier satisfying the equality $ \sum_{i\in\mathcal{I}_j}p_{ji}^{\star} = p_j^{\text{max}} $. Meanwhile, the auxiliaries optimization with fixed power allocation can be obtained through the fixed-point iteration in the form of~(\ref{eq:omega}). The geometry programming can be also exploited to solve the auxiliary optimization, as shown in~\cite{cja}. Finally, the convergence of the alternating optimization processes between power allocation and auxiliaries induces the optimal IoT transmission strategy.

\subsection{Incentive Design}

The incentive design incorporates stake investment and profit-sharing optimization to achieve the highest profit for UAVs in the blockchain network, which is consistent with the objective function in~(\ref{eq:problem}). However, for the practical blockchain system operation, all the concerned parties work in a decentralized manner without explicit outsider coordination. Correspondingly, we intend to design the incentive mechanism from a distributed perspective through game-based analysis, with each concerned party maximizing its own profit.

Without loss of generality, we consider the IoT data to be blockchained originate from cluster-$j$. In accordance with the previous discussions to reduce the system operation overhead, we encourage UAV-$j$ as the currently active data collector working as the miner to generate candidate blocks based on its collected data. To this end, UAV-$j$ constructs a stake pool that allows other UAVs to invest their stakes, while the stake pool shares its obtained payment in return for the investment. In this regard, the pool determines the portion of payment sharing while other UAVs decide their ratio of investment, which leads to a one-leader multi-follower Stackelberg game formulation detailed below.
Specifically, the problem at the leader can be written as
\begin{IEEEeqnarray}{cl}
  \IEEEyesnumber\label{eq:prob_l} \IEEEyessubnumber*
  \max_{\beta_{j}} \quad & \Theta_{j,j} \left(\beta_{j};\bm{\alpha}_{j} \right)  \\
   \rm{s.t.} \quad & \beta_{j}\in[0,1],
\end{IEEEeqnarray}
where $ \bm{\alpha}_j = \left[\alpha_{j,k}\right]_{k\in\mathcal{J}_{-j}} $ is the vector of followers' decision variable, and $ \Theta_{j,j} $ is given in~(\ref{eq:ujj}). Here we explicitly indicate the arguments to show the interplay between different participants in the game. For the followers in the game, the problem is specified as
\begin{IEEEeqnarray}{cl}
  \IEEEyesnumber\label{eq:prob_f} \IEEEyessubnumber*
  \max_{\alpha_{j,k}} \quad & \Theta_{j,k} \left(\alpha_{j,k};\bm{\alpha}_{j,-k}, \beta_j \right)  \\
   \rm{s.t.} \quad & \alpha_{j,k}\in[0,1],
\end{IEEEeqnarray}
for $ k\in\mathcal{J}_{-j} $, where $ \Theta_{j,k} $ is given in~(\ref{eq:ujk}) and $ \bm{\alpha}_{j,-k} = \left[\alpha_{j,k'}\right]_{k'\in\mathcal{J}_{-j}\backslash\{k\}} $ corresponds to all rest followers other than follower-$k$.

 Then, the problem in~(\ref{eq:prob_l}) at the leader and problems in~(\ref{eq:prob_f}) at all followers constitute the Stackelberg game, denoted by $ \mathcal{G} $, which incorporates one leader and $ J-1 $ followers. In the game context, the problem in~(\ref{eq:prob_l}) and problems in~(\ref{eq:prob_f}) are coupled, and thus we cannot directly solve them independently. In particular, for a Stackelberg game, the leader takes action first, followed by the action of the followers, and the solution to the game is defined by the Stackelberg equilibrium. In accordance with the hierarchical structure of decision makings in the Stackelberg game, the equilibrium is also layered. Denote the equilibrium as $ (\beta_j^{\star}, \bm{\alpha}_j^{\star}) $, and then it satisfies the following conditions
\begin{equation} \label{eq:eq_f}
\begin{aligned}
  \Theta_{j,k} \left(\alpha_{j,k}^{\star};\bm{\alpha}_{j,-k}^{\star}, \beta_j^{\star} \right) \ge \Theta_{j,k} \left(\alpha_{j,k};\bm{\alpha}_{j,-k}^{\star}, \beta_j^{\star} \right), \\ \forall \alpha_{j,k} \in [0,1], \quad \forall k\in\mathcal{J}_{-j},
\end{aligned}
\end{equation}
and
\begin{equation} \label{eq:eq_l}
  \Theta_{j,j} \left(\beta_{j}^{\star};\bm{\alpha}_{j}^{\star} \right) \ge \Theta_{j,j} \left(\beta_{j};\bm{\alpha}_{j}^{\star} \right),
\end{equation}
which denote the lower equilibrium for the followers and upper equilibrium for the leader, respectively. Then the lower equilibrium in~(\ref{eq:eq_f}) indicates that no follower will unilaterally deviate from the equilibrium strategy given the leader's action, while the upper equilibrium in~(\ref{eq:eq_l}) implies that the equilibrium strategy at the leader is optimal on condition that the lower equilibrium is also achieved among the followers.

We then find the Stackelberg equilibrium to determine the investment and payment sharing. For Stackelberg games, we employ the backward induction approach to achieve the Stackelberg equilibrium. In this regard, we first analyze the lower equilibrium among the followers while assuming a fixed strategy at the leader, i.e., to find the individually optimal investment with respect to fixed profit sharing. Correspondingly, the followers compete to maximize their own profit function, inducing a generic Nash game model given as
\begin{equation}
  \bar{\mathcal{G}}_{j} (\beta_j) = \left\{ \mathcal{J}_{-j}, [0,1]^{J-1}, \{\Theta_{j,k}\}_{k\in\mathcal{J}_{-j}} \right\},
\end{equation}
where subscript-$j$ specifies the current leader as UAV-$j$ and this game is parameterized by $ \beta_j $ as the leader's strategy. For the Nash game $ \bar{\mathcal{G}}_{j} $, we can easily verify that it is a convex game that the strategy space is compact and convex, and the profit function in~(\ref{eq:ujk}) is continuous against the strategies of all followers and is convex with respect to its own strategy. The proof is omitted here for space limitation while interested readers can refer to the second-order derivative and the conclusion readily follows. Based on the study in~\cite{HanGameBook}, we know that a convex game must admit a Nash equilibrium. To derive the equilibrium, we resort to the best-response strategy, i.e., the optimal strategy of one player on the condition of fixed strategies of others. Given the convexity of the profit function, by nulling the first-order derivative, we can derive the best response for UAV-$k\in\mathcal{J}_{-j}$ given as~(\ref{eq:alpha_star}),
\begin{figure*}
\begin{equation} \label{eq:alpha_star}
  \alpha_{j,k}^{\star} = \left\{
  \begin{array}{cl}
  \min\left\{\frac{1}{\Upsilon_k}\left( \sqrt{\frac{\zeta_j}{\xi_{j,k}} \sum\limits_{k'\in\mathcal{J}_{-j}\backslash\{k\}} \alpha_{j,k'}\Upsilon_{k'} } - \sum\limits_{k'\in\mathcal{J}_{-j}\backslash\{k\}} \alpha_{j,k'}\Upsilon_{k'} \right), 1\right\},&\quad \text{if} \quad \frac{\zeta_j}{\xi_{j,k}} \ge \sum\limits_{k'\in\mathcal{J}_{-j}\backslash\{k\}} \alpha_{j,k'}\Upsilon_{k'} \\
  0, &\quad \text{otherwise}
  \end{array}
  \right. 
\end{equation}
\setcounter{equation}{32}
\begin{equation} \label{eq:extended}
  \alpha_{j,k}^{\star} = \frac{\left(J-2\right)\beta_j\Upsilon_j\left(\Phi + \rho\Psi_j\right)\left[ (1-\beta_j)\left(\Phi + \rho\Psi_j\right) -\sum\limits_{k'\in\mathcal{J}_{-j}}C_{j,k'}  + (J-2)C_{j,k}\right]}{\Upsilon_k\left( (J-1)\left(\Phi + \rho\Psi_j\right)(1-\beta_j) -\sum\limits_{k'\in\mathcal{J}_{-j}}C_{j,k'} \right)^2}
\end{equation}
\setcounter{equation}{27}
\hrulefill
\end{figure*}
where $ \zeta_j $ and $ \xi_{j,k} $ defined as
\begin{equation} \label{eq:aj}
  \zeta_j = \frac{\Upsilon_j}{\sum\limits_{j\in\mathcal{J}}\Upsilon_j} \beta_j\left(\Phi + \rho\Psi_j\right),
\end{equation}
and
\begin{equation} \label{eq:bjk}
  \xi_{j,k} = \frac{1}{\sum\limits_{j\in\mathcal{J}}\Upsilon_j} \left[\left(1-\beta_j\right)\left(\Phi + \rho\Psi_j\right) - \Omega_{j,k}\right],
\end{equation}
are constants obtained by rearranging the terms in the profit functions to simplify the notation.

With the best-response strategy derived in~(\ref{eq:alpha_star}), specifying the optimal strategy for the current followers with respect to the strategies of all other followers, we can then achieve the lower equilibrium through an iterative strategy update process among all followers. However, by revisiting the best-response strategy, we can rearrange the terms in~(\ref{eq:alpha_star}) as
\begin{equation} \label{eq:aux1}
  \frac{\zeta_j\sum\limits_{k'\in\mathcal{J}_{-j}\backslash\{k\}} \alpha_{j,k'}\Upsilon_{k'}}{\left(\sum\limits_{k'\in\mathcal{J}_{-j}} \alpha_{j,k'}\Upsilon_{k'}\right)^2} = \xi_{j,k}.
\end{equation}
By further summing up the equation in~(\ref{eq:aux1}) over all followers, we arrive at
\begin{equation}
  \frac{\zeta_j\left(J-2\right)}{\sum\limits_{k'\in\mathcal{J}_{-j}} \xi_{j,k'}} = \sum\limits_{k'\in\mathcal{J}_{-j}} \alpha_{j,k'}\Upsilon_{k'}.
\end{equation}
The equality above has an implicit assumption that all followers have positive investment at the lower equilibrium. In the case that certain followers are of no investment, then they are regarded as inactive and thus excluded from the competition. Accordingly, the number of remaining active users then replaces the number of followers during the derivation. Then, the equality above is substituted into the equilibrium condition in~(\ref{eq:alpha_star}), and we obtain the following equilibrium
\begin{equation} \label{eq:alpha_close}
  \alpha_{j,k}^{\star} = \frac{1}{\Upsilon_k}\frac{\zeta_j\left(J-2\right)}{\sum\limits_{k'\in\mathcal{J}_{-j}} \xi_{j,k'}}\left( 1- \frac{\xi_{j,k}\left(J-2\right)}{\sum\limits_{k'\in\mathcal{J}_{-j}} \xi_{j,k'}} \right),
\end{equation}
for $ k\in\mathcal{J}_{-j} $, which is further extended as~(\ref{eq:extended}). Finally, the lower equilibrium in~(\ref{eq:extended}) is in closed-form and can be calculated directly without iterations. In this regard, comparing the original equilibrium in~(\ref{eq:alpha_star}) as individual best response requiring iterations, the results in~(\ref{eq:alpha_close}) are achieved due to the special structure of the lower problem and facilitates the calculation of the lower equilibrium.
% Note here the terminology ``closed form'' indicates that no iteration is required for the lower equilibrium among the followers, yet the equilibrium in the form of~(\ref{eq:alpha_close}) is still a function of the upper equilibrium of the leader.
\setcounter{equation}{33}
% \begin{figure*}

% \end{figure*}

With the lower equilibrium obtained in~(\ref{eq:extended}), we can then substitute it into the leader's utility function in~(\ref{eq:ujj}), whose maximization induces the leader's optimal strategy while incorporating the lower equilibrium into consideration. We can adopt one-dimensional search to find the upper equilibrium for the leader, denoted by $ \beta_{j}^{\star} $. We then substitute the leader's strategy $ \beta_{j}^{\star} $ into~(\ref{eq:extended}), leading to the actual lower equilibrium, denoted by $ \bm{\alpha}_j^{\star} \left(\beta_j^{\star}\right) $. Finally, the strategy set $ \left[ \beta_j^{\star}, \bm{\alpha}_j^{\star} \left(\beta_j^{\star}\right) \right] $ constitutes the Stackelberg equilibrium of the incentive game, where $ \beta_j^{\star} $ specifies the portion of profit sharing at UAV-$j$ (the stake pool), and $ \bm{\alpha}_j^{\star} \left(\beta_j^{\star}\right) $ corresponds to the portion of invested stakes at other UAVs. As a further note, based on the deviations above, we can see that the Stackelberg equilibrium uniquely exists, as the leader's problem in~(\ref{eq:prob_l}) admits the optimum, while the followers' strategy is then uniquely determined based on the closed-form expression specified in~(\ref{eq:extended}).

% In this section, we have investigated the IoT transmission and incentive design strategies. We emphasize the the analyses are conducted on one single cluster basis, i.e, exemplified with cluster-$j$. Specifically, we show the transmission strategy between cluster-$j$ and UAV-$j$, and construct a stake pool at UAV-$j$ for the investment from other UAVs, with investment and profit sharing strategies development. The results can be readily extended to other UAV-IoT cluster pairs, allowing their decision-making in a distributed and independent manner.

\section{Outer Problem Solving for UAV Deployment} \label{sec:outer}

In this section, we consider the outer problem for UAV deployment. Similarly, we adopt the distributed decision-making at the UAVs in accordance with the nature of the blockchained IoT system. However, different from the inner problem that can be tackled at each UAV-IoT cluster basis, the UAV deployment affects the blockchain system operation over the whole network, and thus the mutual influence of the strategy at different UAVs needs to be addressed.
% Moreover, in consistence with the overall system metric in~(\ref{eq:problem}), the UAV deployment intends for distributed system optimization, for with the solution of inner problem as discussed in previous section is exploited.

\begin{figure*}[t]
        \centering
        \includegraphics[width=17cm]{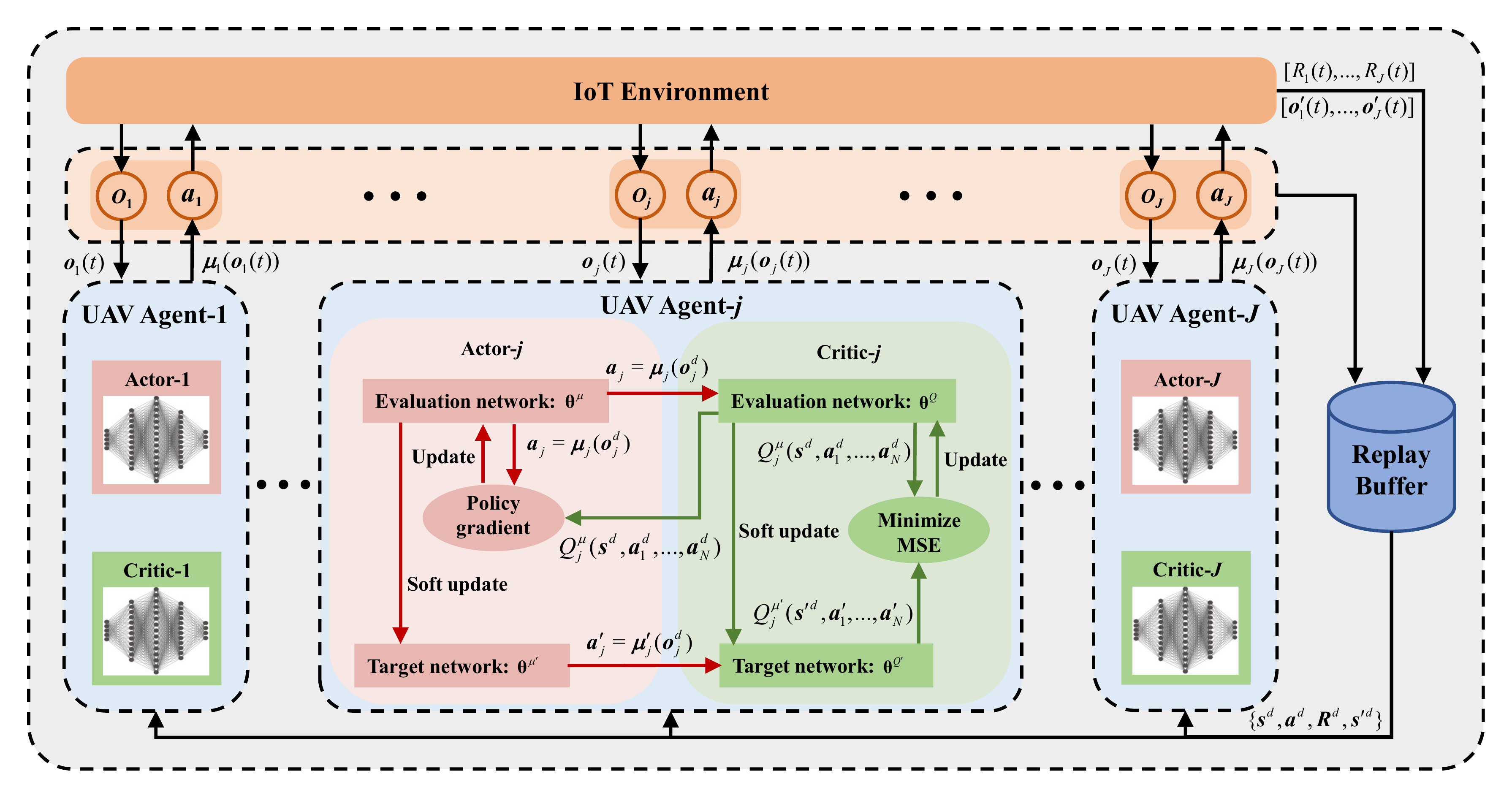}
        \caption{The framework of MADDPG algorithm.}
        \label{fig:maddpg}
\end{figure*}

\subsection{UAV Deployment as a Markov Game}

For certain UAV deployment, it is likely to affect the IoT operations in multiple rounds. Such a relatively long-term effect can be tracked by the Markov decision process (MDP). Meanwhile, each UAV determines the location based on its own observation of the network, while influencing each other. Therefore, the networked problem can be modeled as a partially observable Markov game, which incorporates the deployment of each UAV as a partially observable MDP~\cite{maddpg}. The Markov game with UAVs as $J$ agents can be represented by a set of states $\mathcal{S}$, a set of observations $\mathcal{O}$ = $\left\{\mathcal{O}_1,...,\mathcal{O}_j,...,\mathcal{O}_J\right\}$, a set of actions $\mathcal{A}$ = $\left\{\mathcal{A}_1,...,\mathcal{A}_j,...,\mathcal{A}_J\right\}$, and a reward function for each agent. The state includes the location of IoT nodes and the deployment of UAVs, the observation of an agent is its private version regarding the state, and the action is the decision on an update of deployment. For the current state $\bm{s}\in \mathcal{S}$, each agent-$j$ uses the policy $\bm{\mu}_j: \mathcal{O}_j \mapsto \mathcal{A}_j$ to select an action. Then, the agents interact with the environment by executing the action to achieve a new state, $\mathcal{S} \times \mathcal{A}_1 \times ... \times \mathcal{A}_J \mapsto \mathcal{S}$. In this regard, agent-$j$ receives a reward $R_j: \mathcal{S} \times \mathcal{A}_j \mapsto \mathbb{R}$ according to the state and its own action, along with an updated observation. The goal of each agent is to maximize the long-term expected reward $\Gamma_j=\sum_{t=0}^T\gamma^{t}R_{j}$, where $\gamma\in(0,1)$ is a discount factor, and $t\in[0, T]$ denotes the epoch with $T$ being the time horizon. To be specific, the main elements regarding the formulated Markov game are defined as follows:
\begin{itemize}
  \item State space $\mathcal{S}$: A state $\bm{s}\left(t\right) \in \mathcal{S}$ corresponds to the system environment state at epoch-$t$, including channel state information, and the horizontal coordinates of UAVs and IoT nodes, specified as
  % \begin{equation}
  %   \begin{aligned}
  %     \bm{s}(t) =& \left\{{v}_{1,t}^{(x)},{v}_{2,t}^{(x)},...,{v}_{J,t}^{(x)}, {v}_{1,t}^{(y)},{v}_{2,t}^{(y)},...,{v}_{J,t}^{(y)};\ell_{11,t},\right.\\
  %     &\left.\ell_{12,t},...,\ell_{1I_1,t},...,\ell_{J1,t},\ell_{J2,t},...,\ell_{JI_J,t};{w}_{11,t}^{(x)},\right.\\
  %     &\left.{w}_{12,t}^{(x)},...,{w}_{1I_1,t}^{(x)},...,{w}_{J1,t}^{(x)},{w}_{J2,t}^{(x)},...,{w}_{JI_J,t}^{(x)},{w}_{11,t}^{(y)},\right.\\
  %     &\left.{w}_{12,t}^{(y)},...,{w}_{1I_1,t}^{(y)},...,{w}_{J1,t}^{(y)},{w}_{J2,t}^{(y)},...,{w}_{JI_J,t}^{(y)} \right\}
  %   \end{aligned}
  % \end{equation}
    \begin{equation}
    \begin{aligned}
    \bm{s}\left(t\right) = && \left\{ \left\{\bm{w}_{ji}\left(t\right)\right\}_{i\in\mathcal{I}_j,j\in\mathcal{J}}, \left\{\bm{v}_{j}\left(t\right)\right\}_{j\in\mathcal{J}}, \right. \\
    && \left. \left\{\ell_{ji}\left(t\right)\right\}_{i\in\mathcal{I}_j,j\in\mathcal{J}}  \right\}.
    \end{aligned}
  \end{equation}
  \item Observation space $\mathcal{O}$: Considering that there is no central coordinator for information exchange among UAVs, the UAVs only have the information regarding their own cluster, and thus the observation $\bm{o}_{j}\left(t\right)$ is extended as
  \begin{equation}
    \begin{aligned}
      \bm{o}_{j}\left(t\right) =& \left\{ \left\{\bm{w}_{ji}\left(t\right)\right\}_{i\in\mathcal{I}_j}, \bm{v}_{j}\left(t\right),
    \left\{\ell_{ji}\left(t\right)\right\}_{i\in\mathcal{I}_j}  \right\}.
    \end{aligned}
  \end{equation}
  \item Action space $ \mathcal{A} $: The UAVs take actions to update their deployment. Given current neural network parameter $ \bm{\theta}_{j}^{\mu} $, the policy $\bm{\mu}_{j}(\bm{o}_{j}\left(t\right)|\bm{\theta}_{j}^{\mu})$ induces an actions $\bm{a}_{j}\left(t\right)$ defined as the change on location between two successive epochs, i.e., $ \bm{a}_{j}\left(t\right) = \bm{v}_{j}\left(t\right) - \bm{v}_{j}(t-1)$.
  %  is the change of horizontal coordinates of the UAV positions between the epoch $t$ and epoch $t + 1$, specified as
  % \begin{equation}
  %   a_{j}(t)=\left\{{v'}_{j,t}^{(x)}, {v'}_{j,t}^{(y)} \right\}.
  % \end{equation}
  % respectively, where ${v'}_{j,t}^{(x)}$ and ${v'}_{j,t}^{(y)}$ are the changes in the x-axis and y-axis of the UAV $j$ position, and ${v'}_{j,t}^{(x)}, {v'}_{j,t}^{(y)} \in [-1,1]$.
  \item Reward $R$: The reward measures the effect of the action taken by an agent for a given state, which further guides the agent to find the best deployment strategy. Correspondingly, it needs to be designed in accordance with the system objective. Here we define the reward function as
  \begin{equation}
    R_{j}= U_{j}\left(t\right)-U_{j}\left(t-1\right),
  \end{equation}
  where $ U_j $ is given in~(\ref{eq:uj}) as the achieved utility from the blockchain system given current deployment. In consistence with the definition of action, here we also use the difference between two epochs for the reward function.
  %  Each agent will receive its own reward, which depends on the current state, current action and next state of each agent in the environment $[o_{j}(t), a_{j}(t), o'_{j}(t)]$. The goal of optimization problem is to find the best UAV deployment and transmission power of IoT nodes. In each episode, the agent continuously adjusts actions from the initial state according to different states in different step, so that the UAV can quickly reach the best deployment position, and the cumulative reward is the best. Therefore, the change in the sum utility of all parties verse the time consumed for the whole process between epoch t and epoch $t-1$ is defined as the immediate reward at epoch $t$, given as
  % \begin{equation} \label{eq:y}
  %   \Xi_{j}=\frac{u_{j,j}+\sum_{k\in\mathcal{J}_{-j}}u_{j,k} }{ \tau_j^{(\text{tx})} + \tau_j^{(\text{mn})} + \tau_j^{(\text{pv})} + \tau_{j}^{(\text{cc})} }.
  % \end{equation}
\end{itemize}

\subsection{MADDPG-Based Algorithm}

For problems with high-dimensional state space and continuous action space, the DDPG-based learning can be more effective as compared with conventional approaches such as deep \textit{Q}-learning (DQN) or deterministic policy gradient (DPG). For our considered problem that each agent determines the deployment on their own in a competitive manner based on their local observation, the DDPG approach will be employed at each party involved and thus constitute the MAGGPG framework, which allows independent learning procedure at each UAV while reflecting the interactions with the environment.
% if the state space and action space are discretized, the model is difficult to adapt to high-precision demand scenarios. Therefore, traditional reinforcement Learning algorithms such as Deep Q-learning (DQN) and Deterministic Policy Gradient (DPG) are difficult to quickly and accurately solve the above-mentioned MDP, and it is necessary to use DDPG that can handle high-dimensional state and continuous action space to solve it. In addition, in some environments where there is competition or cooperation between agents, agents often only have partial information about the state of the environment. For each agent, the environment it interacts with will change with the changes of other agents' strategies, which will cause the agent's policy training to fail to converge. Therefore, in order to solve the above problems in the multi-agent environment, a multi-agent reinforcement learning framework is needed to solve the problem of mutual influence between agents through centralized training and decentralized execution.

The proposed MADDPG framework is shown in Fig.~\ref{fig:maddpg}, including a parallel of deep reinforcement learning agents adopting DDPG in an independent while interactive manner. The DDPG learning features an actor-critic architecture, where the actor network based on policy gradient solves the problems with continuous action space, and the critic network based on DQN solves the problems of high-dimensional state space. The actor network determines an action based on the currently observed state and strategy, and the critic network evaluates the action produced by the actor network based on the state-action function. In addition, DDPG also integrates the experience playback of the DQN and the target network, where the experience playback improves the utilization of data and downgrades the correlation of data samples and the target network delays neural network parameter updates and improves the stability of the overall algorithm.

\begin{algorithm} %\small
  \caption{MADDPG for UAV Deployment}
  \label{alg:1}
  \hspace*{0.02in} {\bf Training process:} 
  \begin{algorithmic}[1]
    \STATE {Initialize Critic networks and actor networks for all agents with weights $\left\{\bm{\theta}_{j}^{Q}, \bm{\theta}_{j}^{Q'}, \bm{\theta}_{j}^{\mu}, \bm{\theta}_{j}^{\mu'}\right\}_{j\in \mathcal{J}}$; }
    \STATE {Initialize the reply buffer $\mathcal{D}$;}
    \FOR{episode = 1 to max-episode}
    \STATE{Initialize a random process $\mathcal{N}$ for action exploration;}
    \STATE{Initialize the environment with initial state $\bm{s}_0$;}
    \FOR{$t$ = 1 to max-epoch}
    \STATE{Each agent-$j$ selects an action $\bm{a}_{j}\left(t\right)=\bm{\mu}_{j}\left(\bm{o}_{j}\left(t\right)|\bm{\theta}_{j}^{\mu}\right)+\mathcal{N}$ according to the current neural network with an exploration noise;}
    \STATE{Execute the action $\bm{a}\left(t\right)=\left[\bm{a}_{j}\left(t\right)\right]_{j\in\mathcal{J}}$, obtain reward $\bm{R}\left(t\right)=\left[R_{j}\left(t\right)\right]_{j\in\mathcal{J}}$, and reach a new state $\bm{s}'\left(t\right)$;}
    \STATE{Store transition ($\bm{s}\left(t\right),\bm{a}\left(t\right),\bm{R}\left(t\right),\bm{s}'\left(t\right)$) in $\mathcal{D}$;}
    \STATE{Update the state for the environment;}
    \IF{sufficient transitions collected}
    \FOR{agent $j$ = 1 to $J$}
    \STATE{Sample a random minibatch of $D$ transitions ($\bm{s}^{d},\bm{a}_{j}^{d},R_{j}^{d},\bm{s}'^{d}$) from $\mathcal{D}$;}
    \STATE{Set target \textit{Q}-value according to~(\ref{eq:y})};
    \STATE{Update critic by minimizing the loss in~(\ref{eq:l})};
    \STATE{Update actor using the sampled policy gradient as~(\ref{eq:Del})};
    \ENDFOR
    \STATE{Update target network parameters for each agent based on~(\ref{eq:theta1}) and~(\ref{eq:theta2});}
    \ENDIF
    \ENDFOR
    \ENDFOR
  \end{algorithmic}

\hspace*{0.02in} {\bf Execution process:} 
  \begin{algorithmic}[1]
  \STATE{Load the trained models of critic networks and actor networks of all agents;}
  \STATE{Initialize the environment with initial state $\bm{s}_0$;}
  \FOR{$t $= 1 to max-step}
  \STATE{Each agent selects action according to $\bm{a}_{j}\left(t\right)=\bm{\mu}_{j}\left(\bm{o}_{j}\left(t\right)|\bm{\theta}_{j}^{\mu}\right)$;}
  \STATE{Execute actions $\bm{a}\left(t\right)=\left(\bm{a}_{1}\left(t\right),..,\bm{a}_{J}\left(t\right)\right)$, reach a new state $\bm{s}'\left(t\right)$;}
  \STATE{Update the state of all agents;}
  \ENDFOR
  \STATE{\textbf{Output:} UAV deployment strategy.}
  \end{algorithmic}
\end{algorithm}

The MADDPG-based algorithm design is detailed as Alg.~\ref{alg:1}. The critical procedures are explained as follows.
% For the training process, a critic network that observes the overall situation to guide those with only local information. The actor network is trained and allows the agent to have its own reward function. 
First, initialize all neural network with parameters $\left\{\bm{\theta}_{j}^{\mu}, \bm{\theta}_{j}^{\mu'}, \bm{\theta}_{j}^{Q}, \bm{\theta}_{j}^{Q'}\right\} $ for the evaluation actor network, target actor network, evaluation critic network, target critic network, respectively, at all agent. Also initialize the environment parameters and state before the start of each episode. Secondly, each agent-$j\in\mathcal{J}$ selects and executes actions $ \bm{a}_{j}\left(t\right)$ according to its local observation $\bm{o}_{j}\left(t\right)$, and obtain the immediate reward value $R_{j}\left(t\right)$ with an updated observation $\bm{o}'_{j}\left(t\right)$. Then, store $(\bm{o}_j\left(t\right), \bm{a}_j\left(t\right), {R}_j\left(t\right), \bm{o}'_j\left(t\right))$ for all the agents and constitute a transition in the reply buffer as $(\bm{s}\left(t\right),\bm{a}\left(t\right),\bm{R}\left(t\right),\bm{s}'\left(t\right))$, where $\bm{a}\left(t\right)=[\bm{a}_{j}\left(t\right)]_{j\in\mathcal{J}}$ and $\bm{R}\left(t\right)=[R_{j}\left(t\right)]_{j\in\mathcal{J}}$. After sufficient training, $D$ group of transitions $(\bm{s}^{d}, \bm{a}^{d}, \bm{R}^{d}, \bm{s}'^{d})$ are randomly selected from the experience pool $\mathcal{D}$ for learning. The training goal of the critic network is to reduce the error between the target network parameters and the estimation network parameters, and accurately evaluate the action-value function. Thus the network parameters are updated by minimizing the loss function defined as
\begin{equation} \label{eq:l}
  L\left(\bm{\theta}_{j}^{Q}\right) = \frac{1}{D}\sum_{d=1}^{D}\left(y_{j}^{d}-Q_{j}\left(\bm{s}^{d},\bm{a}^{d}\right)\right)^{2},
\end{equation}
where
\begin{equation} \label{eq:y}
  y_{j}^{d}=R_{j}^{d}+\gamma Q'_{j}\left(\bm{s}'^{d}, \bm{a}'_{1}, ... , \bm{a}'_{J}\right)|_{\bm{a}'_{j}=\bm{\mu}'_{j}\left(\bm{o}_{j}'^{d}\right)},
\end{equation}
$Q_{j}$ and $Q'_{j}$ are action-value functions that take as input the action $\bm{a}$ and the state $\bm{s}$ of all agents, and output the $Q$-value for agent $j$, $D$ is the number of transitions from the minibatch. Then, the actor network maximizes the cumulative reward and updates the network parameters through the gradient ascent method:
\begin{equation} \label{eq:Del}
  \Delta_{\bm{\theta}_{j}^{\bm{\mu}}}J=\frac{1}{D}\sum_{d=1}^{D}\triangledown_{\bm{\theta}_{j}^{\bm{\mu}}}\bm{\mu}\left(\bm{o}_{j}^{d}\right)\triangledown_{\bm{a}_{j}}Q_{j}\left(\bm{x}^{d}, \bm{a}^{d}\right)|_{\bm{a}_{j}=\bm{\mu}_{j}\left(\bm{o}_{j}^{d}\right)}.
\end{equation}
Finally, the MADDPG network parameters are updated as
\begin{equation} \label{eq:theta1}
  \bm{\theta}_{j}^{Q'} \gets \sigma \bm{\theta}_{j}^{Q} + \left(1 - \sigma\right)\bm{\theta}_{j}^{Q'}, \quad \sigma\ll 1,
\end{equation}
and
\begin{equation}  \label{eq:theta2}
  \bm{\theta}_{j}^{\mu'} \gets \sigma \bm{\theta}_{j}^{\mu} + \left(1 - \sigma\right)\bm{\theta}_{j}^{\mu'}, \quad \sigma\ll 1,
\end{equation}
according to the soft update rule, which helps improve the stability of learning.

\section{Simulation Results} \label{sec:sim}

In this section, we present the simulation results to show the performance. We consider an area of 1,000~m $ \times $ 1,000~m. There are 6 IoT clusters while each cluster incorporates 10 IoT nodes and a serving UAV. The nodes are randomly located within the area. The main simulation parameters are summarized in Table~\ref{table}, used as defaults unless otherwise noted.

% In this section, we provide the simulation results to show the performance. We consider a network with 6 UAVs and 60 IoT nodes, forming 6 clusters, with 10 nodes in each cluster. The IoT nodes are uniformly distributed in a square area with a size of 1,000~m $ \times $ 1,000~m. The simulation parameters of the game $\mathcal{G}$ of the blockchain network are: stake $\Upsilon \in [90, 100]$, block packaging fixed reward $\Phi$ = 200, block data processing reward coefficient $\rho$ = 5$\times$10{\textsuperscript{-6}}, operational costs (leader, follow) = (30, 60). Othor simulation parameters are summarized in Table~\ref{table}.

\begin{table}[h]\centering
        \caption{Simulation Parameters}
        \label{table}
        \renewcommand{\arraystretch}{1.5}
        \begin{tabular}{c|c|c}
                % \rowcolor{mygray}
                % \thickhline
                \hline
                \textbf{Parameter} & \textbf{Description} & \textbf{Value} \\
                \hline
                $K$ & Number of antennas at UAV  & 4 \\
                $H$ & Altitude of UAV  & 90 m \\
                $B^{\text{(ag)}}$ & Bandwidth for IoT uplink & 100 kHz \\
                $B^{\text{(aa)}}$ & Bandwidth between UAVs & 100 kHz \\
                $p^{\text{max}}$  & Maximum power per cluster & 1 W \\
                $p^{\text{msk}}$  & Maximum power per node & 0.4 W \\
                $P$  & Transmit power of each UAV & 0.5 W \\
                $\sigma_{0}^2$ & Noise power & -110 dBm \\
                $f$ & Carrier frequency & 2 GHz \\
                $(a, b)$ & Environment factor  & (9.613, 0.158) \\
                $(\eta_{\mathsf{LoS}}, \eta_{\mathsf{NLoS}})$ & LoS and NLoS attenuation  & (1 dB, 20 dB) \\
                % $(\alpha_{\mathsf{LoS}}, \alpha_{\mathsf{NLoS}})$    & Path loss exponent & (2.5, 4.0) \\
                $\tau^{(\text{cc})}$ & Time for confirmation  & 0.5 s \\
                $\Upsilon$ & Available stake at the UAVs  & $\sim U$ (90,100) \\
                $\Psi$ & IoT data size  & 8 M \\
                $\Phi$ & Fixed reward for blockchain  & 200 \\
                $\rho$ & Transaction fee per bit  & 5$\times$10{\textsuperscript{-6}} \\
                $(\Omega_{j,j}, \Omega_{\substack{j,k\\(j\ne k)}}) $ & Blockchain operational cost  & (30, 60) \\
                % $\chi$ & Average size of transactions  & 200 kB \\
                $r_{c}$ & Critic learning rate  & 1$\times$10{\textsuperscript{-5}}  \\
                $r_{a}$ & Actor learning rate  & 1$\times$10{\textsuperscript{-4}}  \\ 
                $\gamma$ & Discounted factor  & 0.90  \\
                min-batch & Batch size  & 128  \\     
                $\left| \mathcal{D} \right|$ & Buffer capacity  & 10{\textsuperscript{6}}  \\
                max-step & Maximum epoch  & 250  \\
                max-episode & Maximum episode  & 5000  \\
                \hline        
        \end{tabular} 
\end{table}

\subsection{Convergence of MADDPG and UAV Deployment}

\begin{figure}[t]
        \centering
        \includegraphics[width=7.8cm]{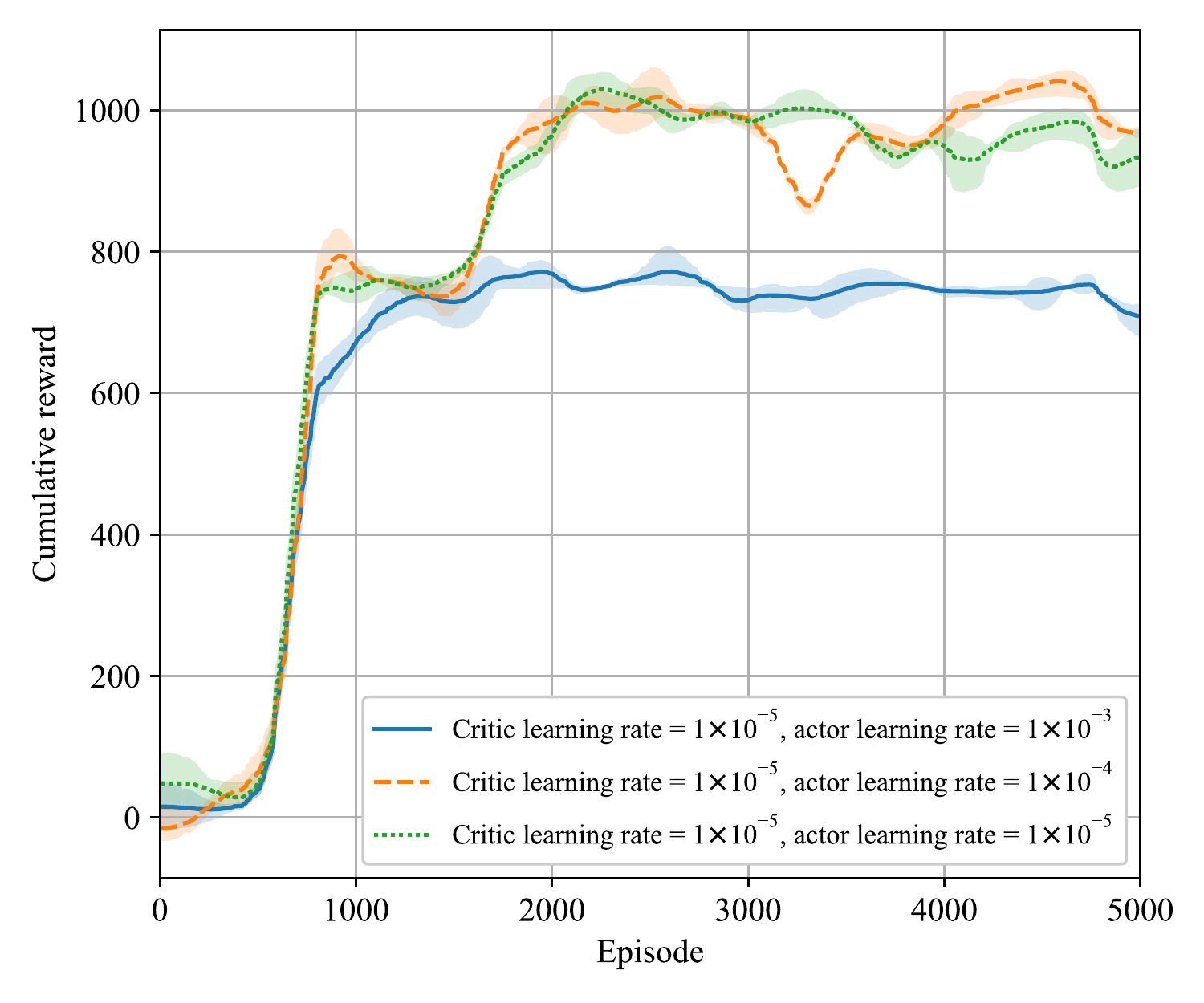}
        \caption{Cumulative reward with different learning rate of the actor network.}
        \label{fig:itera}
\end{figure}

\begin{figure}[t]
        \centering
        \includegraphics[width=7.8cm]{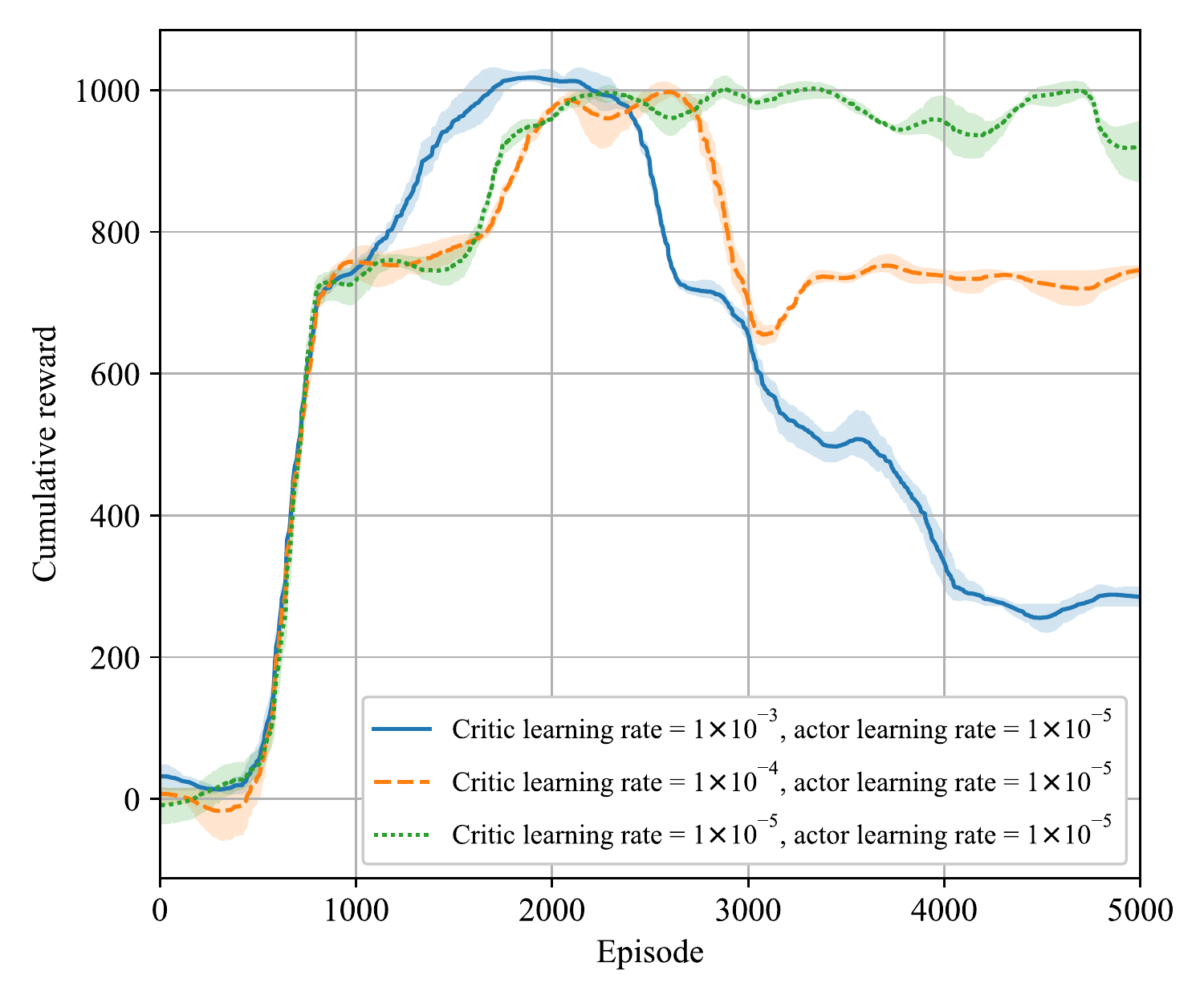}
        \caption{Cumulative reward with different learning rate of the critic network.}
        \label{fig:iterc}
\end{figure}

We first show the convergence of the proposed MADDPG approach in Figs.~\ref{fig:itera} and~\ref{fig:iterc}, where the cumulative reward under the MADDPG algorithm is shown with different actor network learning rates and different critic network learning rates. It can be readily seen that the training process eventually converges, while the learning rate has a significant impact on the convergence rate. During the training process, if the learning rate is too large, it is likely to induce an overfitted neural network after training, and thus the cumulative reward will fluctuate or may even diverge. On the contrary, if the learning rate is too small, it leads to slow convergence during the training process. Therefore, setting an appropriate learning rate is crucial for actual algorithm implementation. Also, it should be noted that with proper learning rates, the proposed approach can solve our formulated problem effectively.

% \begin{figure}[t]
%         \centering
%         \includegraphics[width=7.5cm]{loss_a_c.eps}
%         \caption{Loss function iterations of all agents}
%         \label{fig:iterloss}
% \end{figure}

% In Fig.~\ref{fig:iterloss}, we shows the convergence process in a micro-scope with the value of loss function at each agent's actor network and critic network. Here, we show agent-1,3,5 for a clear demonstration while omitting the results for agent-2,4,6. Among them, the loss function of the actor network is the negative number of the cumulative reward expectation of the strategy. As the number of training steps increases, the parameters of the actor network are continuously optimized, and the loss value is continuously reduced until convergence. The loss function of the critic network is the error between the target network and the estimated network. The smaller the error, the better the optimization of the parameters of the critic network.

% \subsection{UAV Deployment and Profit}

\begin{figure}[t] %\vspace{1pt}
        \centering
        \includegraphics[width=7.55cm]{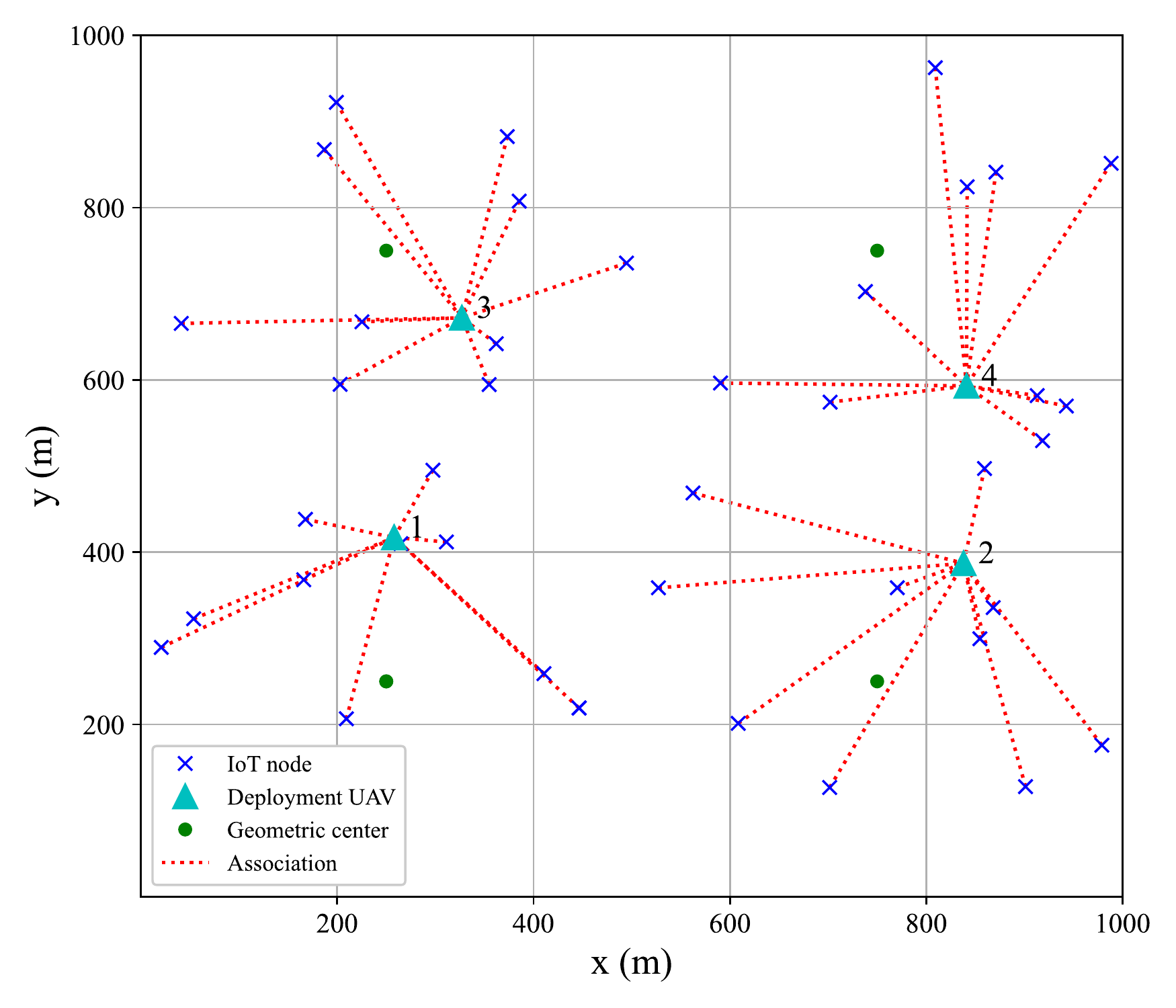}
        \caption{An illustration of UAV deployment with 4 IoT clusters.}
        \label{fig:topo4}
\end{figure}
\begin{figure}[t] %\vspace{pt}
  \centering
  \includegraphics[width=7.55cm]{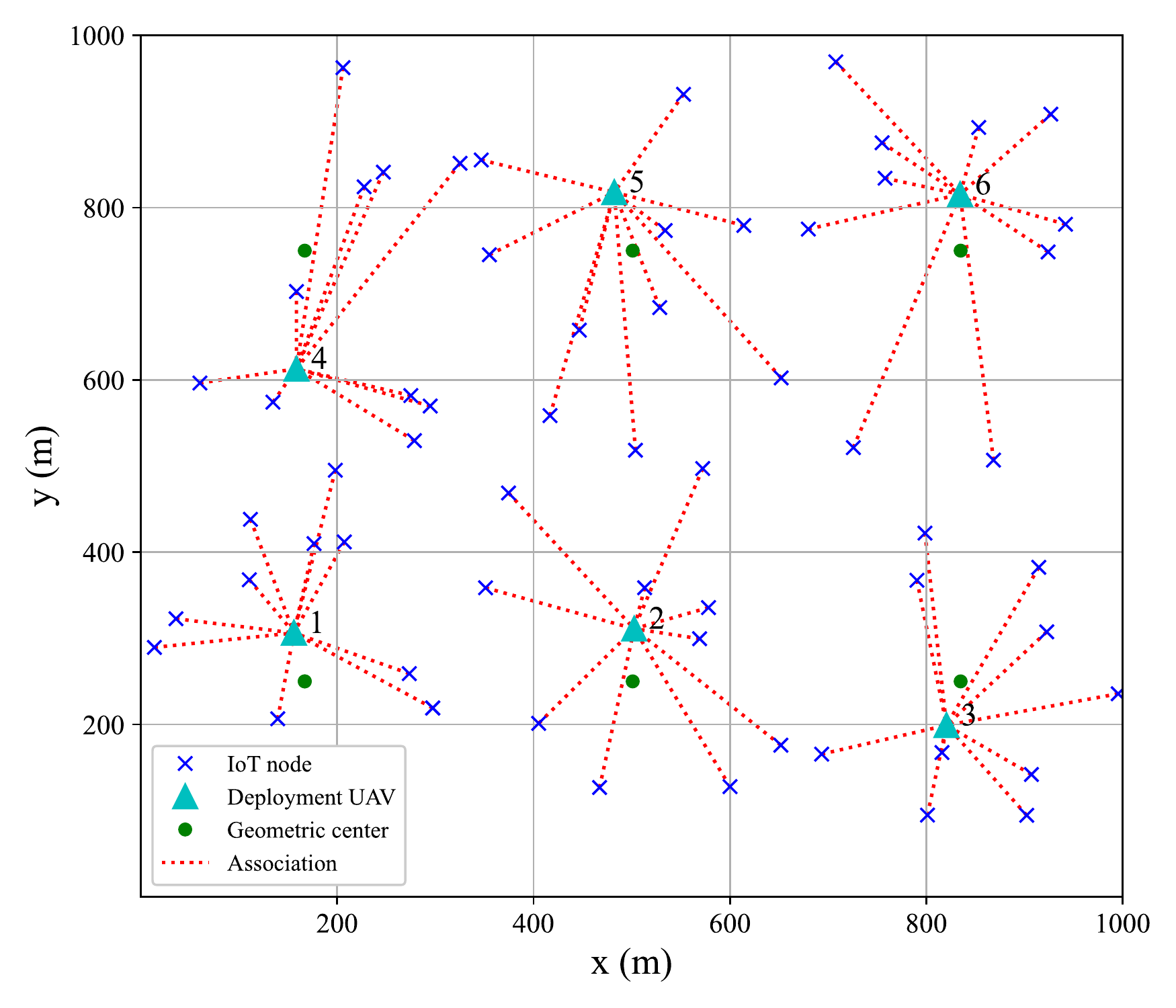}
  \caption{An illustration of UAV deployment with 6 IoT clusters.}
  \label{fig:topo6}
\end{figure}

\begin{figure}[t] %\vspace{2pt}
        \centering
        \includegraphics[width=8.7cm]{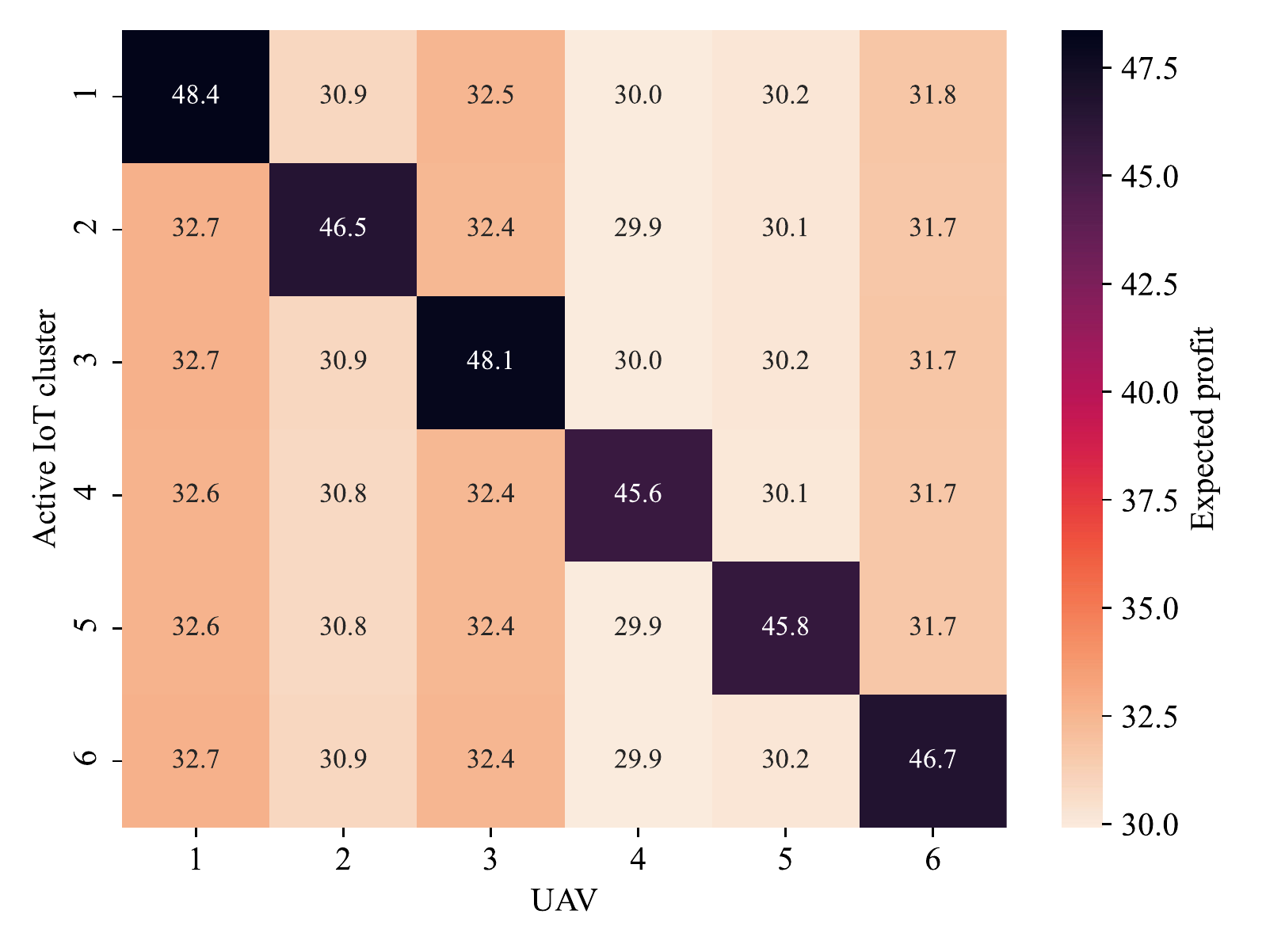}
        \caption{Expected profit at the UAVs with incentive design.}
        \label{fig:Heat}
\end{figure}

In Figs.~\ref{fig:topo4} and~\ref{fig:topo6}, we show the UAV deployment through the proposed MADDPG approach with different network settings and topology, where the cases with 4 clusters and 6 clusters are shown. We also explicitly indicate the geographic centers for the areas for each cluster. The results depict two trends regarding the UAV deployment for the blockchain-secured IoT. The UAV tends to hover at the location with more nodes gathering. In this regard, the UAV can establish the IoT uplink with improved channel quality so as to facilitate the data collection with reduced transmission time and improve the system utility. Meanwhile, the UAVs as blockchain users, tend to locate close to each other. In this regard, the UAVs have closer air transmission distances, for which the block propagation and verification can be more conveniently conducted, further reducing the blockchain operation latency and improving the system utility. While these two trends sometimes coincide, as shown in Fig.~\ref{fig:topo4}, sometimes compromise is needed, as shown in Fig.~\ref{fig:topo6}, which reveals that the multiple factors affecting the UAV deployment need to be well balanced for the optimized system performance.

% strategies under different topologies optimized by the MADDPG algorithm, where the IoT node locations are randomly generated in the deployment area. It can be seen that for each cluster, the UAV is located near the center of the node gathering area in the cluster to balance the transmission quality of the nodes in the IoT cluster. In addition, due to the need for the block propagation and verification process of communication between drones, drones tend to be close to each other to reduce the time delay in the PoS process to obtain higher system performance.

\subsection{Performance Comparison}

In Fig.~\ref{fig:Heat}, we show the obtained profit through the proposed incentive mechanism. In particular, each row corresponds to the case with one active IoT cluster. For example, for the first row, cluster-1 is now actively conducting IoT transmission and thus UAV-1 constructs a stake pool allowing the investment from others. In this regard, we can see that at the Stackelberg equilibrium, UAV-1 with the pool obtains a profit of 48.4 coins, while other UAVs have a profit at about 30 coins. Similar results can be observed in cases with different active IoT clusters and pool constructions. Note that the Stackelberg game with leader-follower architecture allows the leader to take action first and achieve an advantageous position in the game, and the pool has a relatively higher profit as compared with the followers. Note for the case without pools, all UAVs have an equal position in the game and thus there are no leader's advantages. Since the expected profit considers the cases for other UAVs working as the miner to get the payment for blockchain operation as well as the reward for investment. In this regard, even the pool is constructed, the mining process does not necessarily occur here. However, the pool construction does induce a higher expected profit, even though the pool needs to share its payment for the investment. Therefore, we can see that the proposed investment mechanism is quite effective in improving the utility at the pool constructor with reduced cost, which further improves the system utility of the overall blockchain system.

 % compare the profitability of all participants in the stake pools of different drones in the equilibrium of the blockchain game. It can be seen that under the optimal response conditions, the leader's profit in the stake pool is relatively larger than that of the follower, which is affected by the difference in operational costs between the leader and the follower and the investment proportion of the follower. Since the transmission and storage of data is time-consuming and resource-intensive, the leader's production of blocks requires less overhead than the follower's production. In order to maximize the profits of the entire stake pool, followers tend to invest their own shares in the leader and share the profits of the stake pool in proportion to the investment. This encourages more network nodes to join the stake pool, thereby improving network performance and security.

% \begin{figure}[t]
%         \centering
%         \includegraphics[width=9cm]{ALL_U_JK.eps}
%         \caption{Pool's profit from the leader and followers}
%         \label{fig:3D}
% \end{figure}

\begin{figure}[t] \vspace{-2pt}
        \centering
        \includegraphics[width=7.8cm]{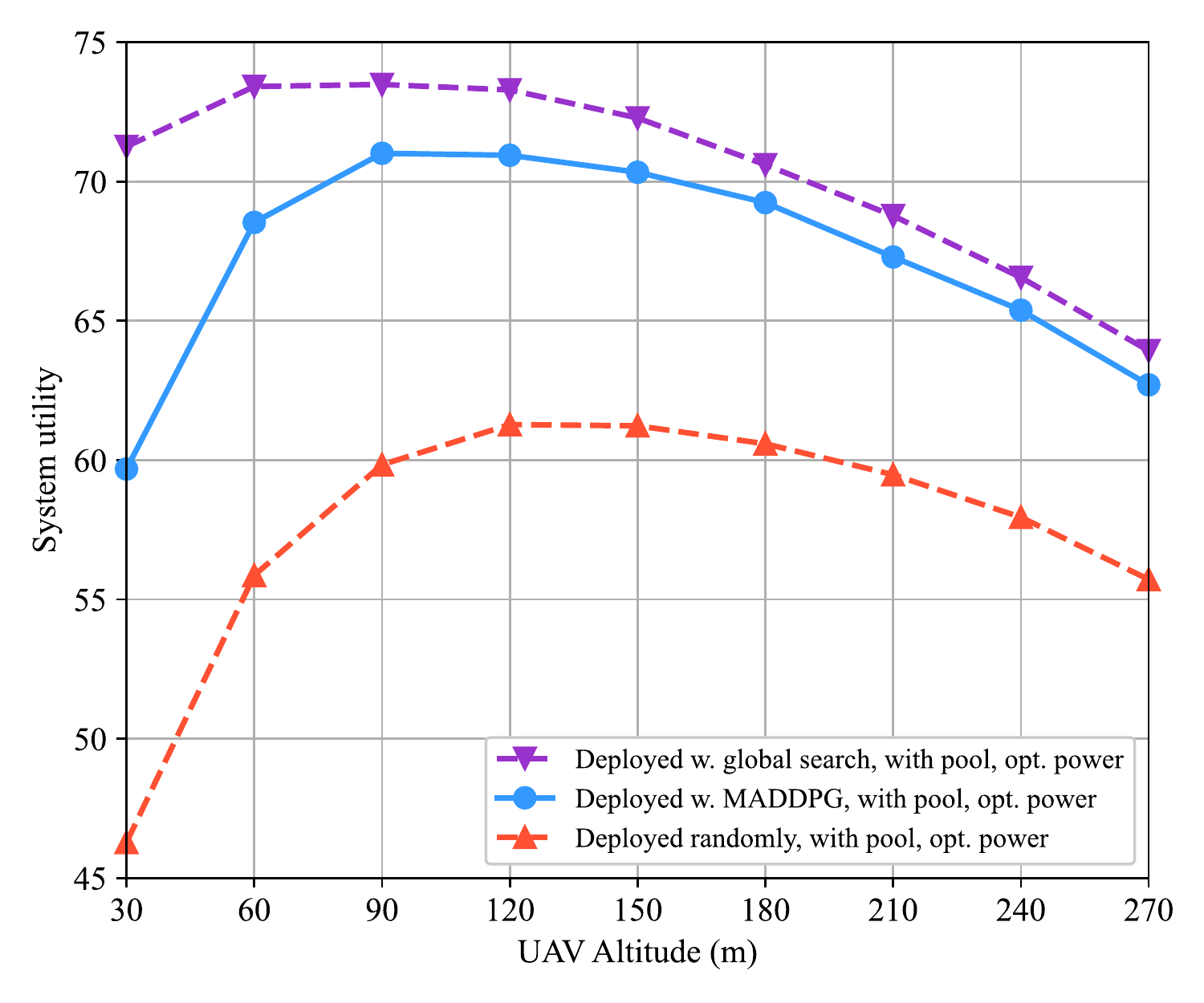}
        \caption{Performance comparison with different approaches.}
        \label{fig:alg}
\end{figure}

In Fig.~\ref{fig:alg}, we show the results comparing the proposed MADDPG approach with global search as well as the results with random UAV deployment as baselines. As we mainly address the deployment through MADDPG, and thus we adopt the IoT transmit power allocation and incentive design as analyzed in Sec.~\ref{sec:inner} to facilitate the discussion, and use UAV altitude as the ranking variable. For all considered approaches, the achieved system utility first increases and then decreases as the UAVs reach higher. This can be explained by the IoT-UAV transmission link quality that first improves and then downgrades, inducing a larger probability for LoS transmissions while worsened channel quality as the UAV altitude increases. Thus, there exists a tradeoff in terms of UAV altitude for the system performance. Meanwhile, as expected, the global search provides the best performance, while our proposed MADDPG can effectively approach the optimum, while there is an evident performance gap with random UAV deployment as compared with other approaches. Particularly, we can see that our proposal more closely approximates the optimum when the UAVs fly higher. This is because, with higher UAV altitude, the location of IoT nodes in each cluster has a smaller impact over the UAV deployment, since the transmission link from the UAV to the IoT nodes tends to be the same. In this regard, the UAV deployment will emphasize more on the blockchain system and thus locate closer to each other to facilitate the blockchain operations. Therefore, the results for UAV deployment tend to be the same even different approaches are exploited. Moreover, we emphasize that the global search as a centralized approach requires explicit coordination among different UAVs and clusters, which goes against the distributed nature of IoT and blockchain system. In contrast, our proposed MADDPG approach can be implemented in a distributed manner, which is thus not only effective but also convenient.

% In Fig.~\ref{fig:alg}, we compare the global search algorithm with MADDPG, DDPG and random algorithms to optimize the system performance of UAV deployment. It can be seen that the MADDPG algorithm obtains better system performance than the random algorithm, and the optimization effect is close to the global optimum, which proves the effectiveness of our proposed algorithm. In addition, although MADDPG is slightly inferior to the single agent-based DDPG algorithm in terms of optimization results, in the application of the DDPG algorithm, a central controller is required to provide information exchange between drones, which requires additional spectrum and time costs. In most multi-agent scenarios, MADDPG-based solutions can save costs and adapt to environmental changes better than DDPG-based solutions.

\begin{figure}[t] \vspace{-5pt}
        \centering
        \includegraphics[width=7.8cm]{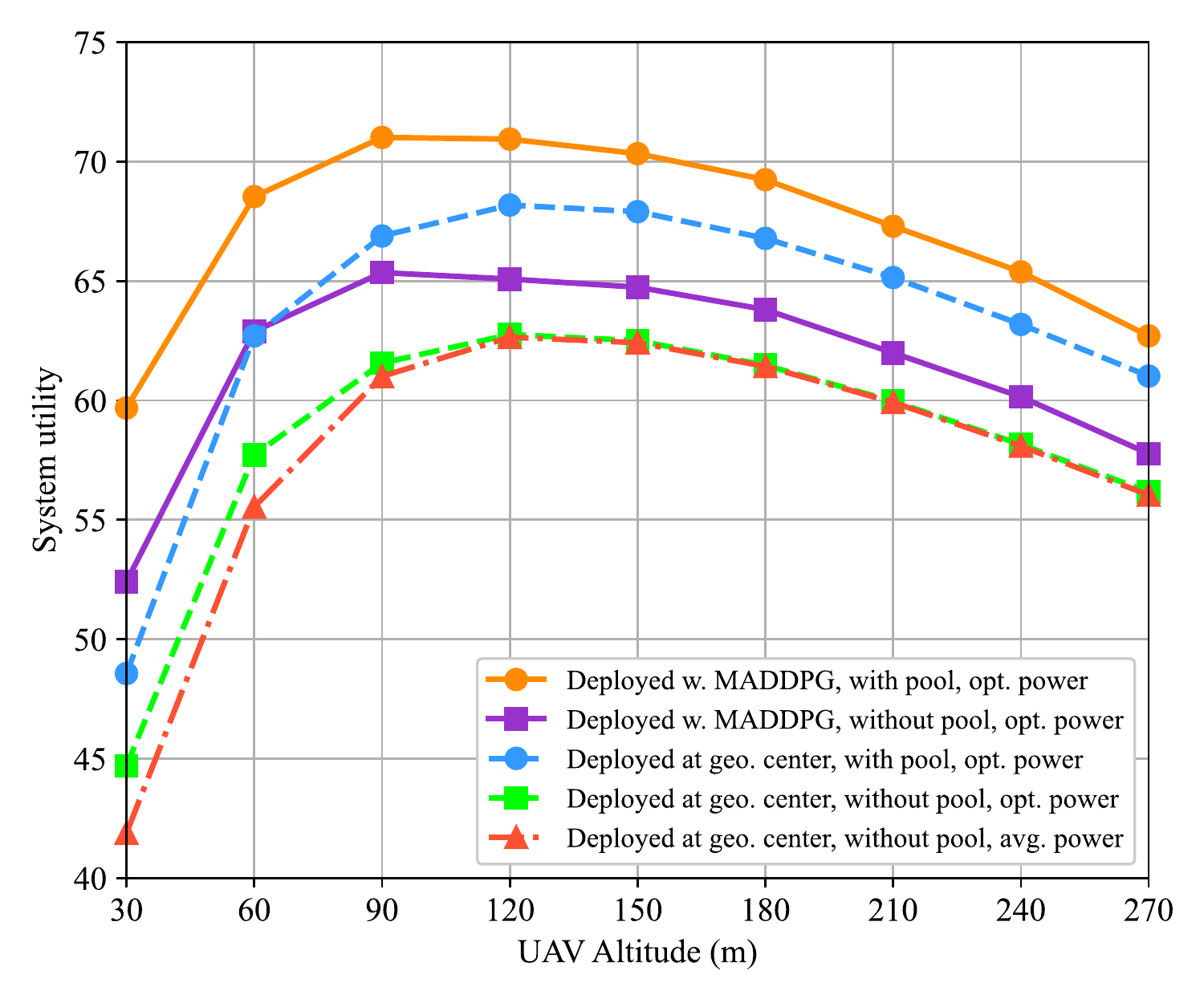}
        \caption{System utility with respect to UAV altitude.}
        \label{fig:PPP}
\end{figure}

In Fig.~\ref{fig:PPP}, we compare the performance under our proposal with the cases with the different transmission, incentive, and deployment strategies, as specified in the legend of the figure. The performance indicator is the system utility defined as the objective function in~(\ref{eq:problem}). Similar to the results in Fig.~\ref{fig:alg}, the performance first improves and then downgrades as the UAV altitude increases, due to the tradeoff between LoS probability and channel propagation for the air-ground transmissions. Meanwhile, our proposal results in the best performance as compared with the baselines, regardless of the UAV altitude. Comparing the results under different approaches, we can see the performance gain through the learning-based deployment (solid curves vs dashed curves), incentive design (curves with circle markers and curves with square markers), and the IoT transmission optimization (curves with triangle markers and the others). Overall, as the UAV becomes higher, the link quality from different IoT nodes to the UAV becomes closer, and thus the performance gap among different approaches shrinks. An interesting observation is that, as we compare the results from MADDPG without the pool and geographic center deployment with pool constructions, we can see that the former outperforms the latter when the UAV altitude is relatively low. This is because, when the UAV is relatively lower, the difference in channel quality from different IoT nodes to the UAV becomes more evident due to the stronger NLoS components. In this regard, the UAV deployment can more significantly influence the overall performance as deployment through learning is generally superior to the geographic center deployment. Then, the performance gain from deployment optimization outweighs that from the pool construction. In contrast, when the UAV altitude becomes higher, the difference in IoT transmission links in the same cluster becomes smaller. In this regard, the deployment through MADDPG approximates the geographic center deployment. Meanwhile, the benefit brought by pool construction becomes more evident due to the improved blockchain profit and reduced cost.

% As expected, our proposal outperforms the baseline scheme with or without unilateral optimization. It can be seen that as the height of the UAV increases from 30 m to 270 m, the performance of the system first increases and then decreases. This is due to the nature of the air-to-ground communication channel, a lower height will increase the probability of non-line-of-sight transmission but will shorten the transmission distance, and vice versa. Therefore, the communication distance and line-of-sight transmission probability make up for each other, for which there is an optimal deployment height for UAVs. 

% UAV deployment plays a more important role in overall performance, because it not only affects IoT data collection, but also affects the PoS process delay in blockchain operations. In addition, the design of the stake pool increases the average profit obtained by blockchain participants, thereby attracting more stakeholders to participate in investment, and improving the security performance of the blockchain network. In general, we can see that the optimization of resources and topology is essential for the efficient operation of the blockchain in the context of the wireless Internet of Things.

\begin{figure}[t] \vspace{-5pt}
        \centering
        \includegraphics[width=7.8cm]{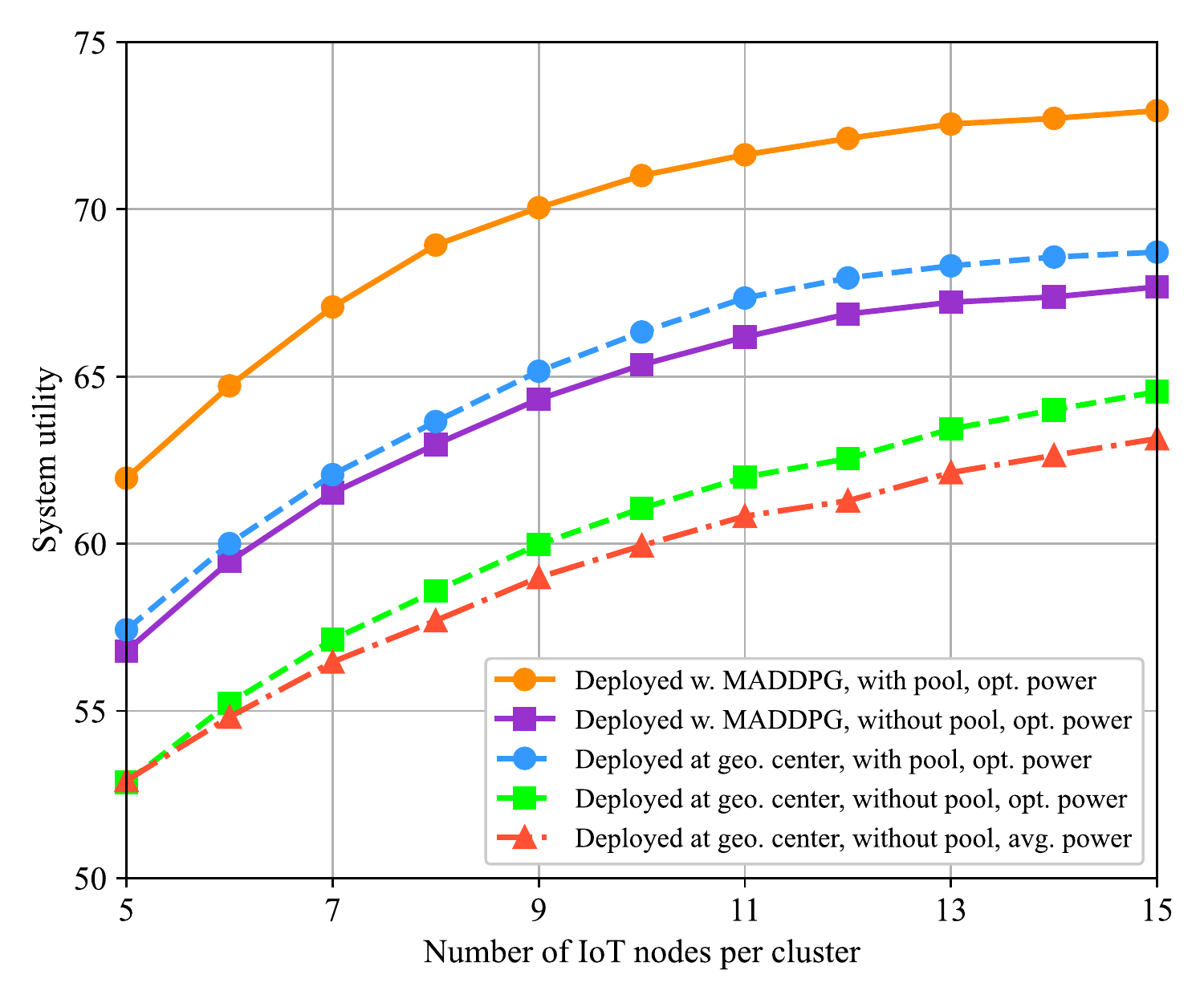}
        \caption{System utility with respect to number of nodes in a cluster.}
        \label{fig:uer}
\end{figure}

Fig.~\ref{fig:uer} shows the performance comparison considering the number of IoT nodes per cluster, where the baselines are similarly defined as those in Fig.~\ref{fig:PPP}. Also similarly, we can see that our proposal with joint deployment, incentive, and transmission optimization outperforms the rest. Also, the system utility is improved with an increasing number of nodes in the IoT cluster for all approaches. This is as expected since more IoT nodes induce improved IoT uplink transmissions to facilitate the blockchain operations. For the results from different approaches, we observe the performance gained by different factors. In particular, the design of the stake pool increases the average profit obtained by blockchain participants, thereby attracting more stakeholders to participate in the investment, and improving the performance of the blockchain network. Further, the UAV deployment and pool construction both have significant impact on the overall performance, since the deployment affects the IoT transmission as well as the blockchain operations while pool construction direct affects the profit. Moreover, with more IoT nodes, the performance gap between different approaches becomes larger. This is because, where there are more nodes, the transmission link difference between IoT nodes becomes more significant, which allows a larger space for the optimization in different aspects, including the deployment, incentive, and transmissions, to demonstrate their effectiveness and superiority.

% For the cases with and without pool construction, we can see that the incentivized investment with profit sharing helps improve the system performance with reduced operational cost. Moreover, we can see an interesting fact that for the cases of MADDPG deployment without pools, and geographic-center deployment with pool construction, the former results in higher system utility with less IoT nodes, while the contrary holds for the latter case. This indicates that when the IoT nodes are sparely distributed, the difference between different deployment strategies is rather significant. While with more nodes, the deployment from different strategies tends to be more close, therefore allowing a more evident performance difference regarding pool construction.

% In addition, due to the limitation of network coverage, the system performance advantage of MADDPG-based deployment as compared with geometric center deployment decreases as the number of nodes increases. Therefore, as the number of nodes increases, the system performance curve of the MADDPG without stake pool strategy and the geometric center deployment with stake pool strategy will intersect.

% \newpage

\section{Conclusion} \label{sec:con}

In this paper, we proposed a UAV-assisted data collection for clustered IoT with PoS blockchain-based security, where the IoT transmission, incentive for PoS, and UAV deployment were jointly considered. The proposed MADDPG was exploited for the UAV-IoT cluster pair to learn their strategy, facilitating the implementation in a distributed manner. We particularly proposed the incentive design for the PoS procedure with stake pool construction allowing investment-based profit sharing. The numerical results indicated that implicit coordination is required for IoT transmission and blockchain operation in terms of UAV deployment for the optimized system performance. Moreover, the construction of a stake pool not only facilitated the PoS consensus procedure for blockchain, but also worked as an effective incentive mechanism to improve the overall system utility.

% In this paper, we exploit the PoS-based blockchain technology to secure the data collection for UAV-assisted IoT, where a DDPG-based approach is proposed to solve the joint IoT transmission and UAV deployment issues. With verified the convergence of the proposed scheme, the numerical results indicate that the there are tradeoffs in terms of UAV altitude the deployment for the system performance. Moreover, the coordination between IoT communication and blockchain operation is critical to improve the blockchain throughput as the transmission delay and block propagation delay need to be well balanced.

\bibliographystyle{IEEEtran}
\bibliography{mainBib}

\end{document}